\documentclass[a4paper,11pt]{article}	
\pdfoutput=1
\usepackage{dcolumn}

\usepackage{bm}
\usepackage[T1]{fontenc}
\usepackage{graphicx}
\usepackage{amssymb,amsmath}
\usepackage{booktabs}
\usepackage{multirow}
\usepackage{cite,color,url}
\usepackage[dvipsnames]{xcolor}
\def\linkcolor{green!70!black}

\usepackage[
colorlinks=true
,urlcolor=\linkcolor
,anchorcolor=\linkcolor
,citecolor=\linkcolor
,filecolor=\linkcolor
,linkcolor=\linkcolor
,menucolor=\linkcolor
,linktocpage=true
,pdfproducer=medialab
,pdfa=true
]{hyperref}

\usepackage{slashed}
\usepackage{epsfig,psfrag,rotating,soul}
\usepackage{rotfloat}
\usepackage[font={small}]{caption}
\usepackage{colortbl}
\usepackage{soul}
\usepackage{xtab}
\usepackage{tikz}
\usepackage{makecell}

\evensidemargin \oddsidemargin
\newcommand{\be}{\begin{equation}}
\newcommand{\ee}{\end{equation}}

\newcommand{\beq}{\begin{equation}} 
\newcommand{\eeq}{\end{equation}} 
\newcommand{\ba}{\begin{array}}  
\newcommand{\ea}{\end{array}} 
\newcommand{\bea}{\begin{eqnarray}}  
\newcommand{\eea}{\end{eqnarray} }  
\newcommand{\bal}{\begin{align}}
\newcommand{\eal}{\end{align}}   
\newcommand{\bi}{\begin{itemize}}  
\newcommand{\ei}{\end{itemize}}  
\newcommand{\ben}{\begin{enumerate}}
\newcommand{\een}{\end{enumerate}}  
\newcommand{\bc}{\begin{center}}
\newcommand{\ec}{\end{center}} 
\newcommand{\bt}{\begin{table}}
\newcommand{\et}{\end{table}}  
\newcommand{\btb}{\begin{tabular}}
\newcommand{\etb}{\end{tabular}}

\allowdisplaybreaks

\let\OLDthebibliography\thebibliography
\renewcommand\thebibliography[1]{
  \OLDthebibliography{#1}
  \setlength{\parskip}{0pt}
  \setlength{\itemsep}{0pt plus 0.3ex}
}

\usepackage{fontawesome5}
\makeatletter
\newcommand{\github}[1]{%
   \href{#1}{\faGithubSquare}%
}
\makeatother

\begin{document}

\begin{titlepage}

\thispagestyle{empty}

\def\thefootnote{\fnsymbol{footnote}}

\begin{flushright}
EFI-24-9\\
IFT-UAM/CSIC-24-91\\
\end{flushright}

\vspace*{1cm}

\begin{center}

\begin{Large}
\textbf{
Machine-Learning Analysis of Radiative Decays \\
to Dark Matter at the LHC
}
\end{Large}

\vspace{1cm}

{\sc
Ernesto~Arganda$^{1}$%
\footnote{{\tt \href{mailto:ernesto.arganda@uam.es}{ernesto.arganda@uam.es}}}%
, Marcela~Carena$^{2, 3, 4}$%
\footnote{{\tt \href{mailto:carena@fnal.gov}{carena@fnal.gov}}}%
,  Mart\'in~de~los~Rios$^{1, 5}$%
\footnote{\tt \href{mailto:martindelosri@sissa.it}{mdelosri@sissa.it}}%
, Andres D. Perez$^{1}$%
\footnote{\tt \href{mailto:andresd.perez@uam.es}{andresd.perez@uam.es}}%
, Duncan~Rocha$^{2, 3}$%
\footnote{\tt \href{mailto:drocha@uchicago.edu}{drocha@uchicago.edu}}%
, Rosa~M.~Sand\'a Seoane$^{1}$%
\footnote{{\tt \href{mailto:rosa.sanda@uam.es}{rosa.sanda@uam.es}}}%
and Carlos~E.~M.~Wagner$^{3, 4, 6}$%
\footnote{{\tt \href{mailto:cwagner@uchicago.edu}{cwagner@uchicago.edu}}}%
}

\vspace{0.5truecm}

{\sl
$^1$Departamento de Física Teórica and Instituto de F\'{\i}sica Te\'orica UAM-CSIC, \\
Universidad Autónoma de Madrid, Cantoblanco, 28049 Madrid, Spain

\vspace*{0.15cm}

$^2$Fermi National Accelerator Laboratory, P. O. Box 500, Batavia, IL 60510, USA

\vspace*{0.15cm}

$^3$Enrico Fermi Institute, Physics Department, University of Chicago, Chicago, IL 60637, USA

\vspace*{0.15cm}

$^4$Kavli Institute for Cosmological Physics, University of Chicago, Chicago, IL 60637, USA

\vspace*{0.15cm}

$^5$SISSA -  International School for Advanced Studies, Via Bonomea 265, 34136 Trieste, Italy

\vspace*{0.15cm}

$^6$HEP Division, Argonne National Laboratory, 9700 Cass Ave., Argonne, IL 60439, USA
}

\vspace*{2mm}

\end{center}

\vspace{0.1cm}

\renewcommand*{\thefootnote}{\arabic{footnote}}
\setcounter{footnote}{0}

\begin{abstract}
\noindent 
The search for weakly interacting matter particles (WIMPs) is one of the main objectives of the High Luminosity Large Hadron Collider (HL-LHC). In this work we use Machine-Learning (ML) techniques to explore 
WIMP radiative decays into a Dark Matter (DM) candidate in a supersymmetric  framework. The minimal supersymmetric WIMP sector includes the lightest neutralino that can provide the observed DM relic density through its co-annihilation with the second lightest neutralino and lightest chargino.
Moreover, the direct DM detection cross section rates fulfill current experimental bounds and provide discovery targets for the same region of model parameters in which the radiative decay of the second lightest neutralino into a photon and the lightest neutralino is enhanced. This strongly  motivates the search for radiatively decaying neutralinos which, however, suffers from strong backgrounds.  We investigate the LHC reach in the search for these radiatively decaying particles by means of cut-based and ML methods and estimate its discovery potential in this well-motivated, new physics scenario. We demonstrate that using ML techniques would enable access to most of the parameter space unexplored by other searches.

\end{abstract}

\end{titlepage}

\tableofcontents

\section{Introduction}\label{sec:intro}

The origin of the scale of electroweak (EW) symmetry breaking and the identity of the constituent particles that make up Dark Matter (DM) are two of the big questions in particle physics. Beyond the Standard Model (BSM) theories with a weakly interacting massive particle (WIMP)~\cite{Steigman:1984ac,Arcadi:2017kky} sector provide  a natural framework that connects these two important questions and has been the subject of intense exploration at colliders and direct detection experiments for the past few decades. The exploration of a WIMP sector as a solution to the DM mystery,  together with a deeper understanding of the Higgs boson properties~\cite{ParticleDataGroup:2024cfk}, are the two main objectives of the Large Hadron Collider (LHC) and its upcoming 
High Luminosity (HL) run. The HL-LHC will provide an unprecedented amount of data, opening new windows into DM searches, yet posing new challenges for  disentangling potential new physics signals from a plethora of background signals. 

Supersymmetry~\cite{Fayet:1976et,Fayet:1977yc,Nilles:1983ge,Haber:1984rc,Gunion:1984yn,Martin:1997ns} has been and remains a leading BSM theory to
connect the two big questions: i) it readily provides a DM candidate by rendering the  lightest supersymmetric particle (LSP) stable in the presence of $R$-Parity~\cite{Farrar:1978xj,Dimopoulos:1981zb,Weinberg:1981wj,Sakai:1981pk,Dimopoulos:1981dw}; ii) it relates the weak scale to the soft supersymmetry breaking scale by naturally explaining the smallness of the measured Higgs boson mass with respect to the Planck scale of new physics.
Current searches at the LHC~\cite{ATLAS:2017tmw,ATLAS:2018nud,CMS:2019zmd,ATLAS:2020syg,ATLAS:2021twp,ATLAS:2021kxv,CMS:2021eha,CMS:2023yzg,ATLAS:2023xco,ATLAS:2024lda} provide significant  constraints on the presence of weak scale supersymmetric partners of the colored Standard Model (SM) particles.  WIMPs, instead, may be lighter~\cite{ATLAS:2024lda,CMS:2024gyw,ATLAS:2024qxh}, and due to their small production cross sections, their potential weak scale masses can be  efficiently probed  at the HL-LHC runs.

There is currently only indirect information about the characteristic scale of potential supersymmetric weakly interacting  particles. This comes primarily  from cosmological requirements to produce the observed dark matter relic density~\cite{Planck:2018vyg}.  A proper contribution to the DM relic density may be achieved in the co-annihilation region~\cite{Ellis:1998kh,Ellis:1999mm,Buckley:2013sca,Han:2013gba,Cabrera:2016wwr,Baker:2018uox,Yanagida:2019evh,Baer:2005jq}, where the mass of the lightest neutralino is close to other weakly interacting particles with higher annihilation rate.  A particularly interesting example that arises somewhat naturally in supersymmetric scenarios is when the LSP is the superpartner of the hypercharge gauge boson (bino), $\tilde \chi_1^0$, and these additional particles are the second lightest neutralino, $\tilde \chi_2^0$, and the lightest chargino, $\tilde \chi_1^\pm$, which in the Minimal Supersymmetric Standard Model (MSSM)~\cite{Fayet:1976et,Fayet:1977yc,Nilles:1983ge,Haber:1984rc,Gunion:1984yn,Martin:1997ns} can be, for instance, the superpartners of the $Z$ and $W$ gauge bosons (winos), respectively. The mass of the LSP  is constrained by direct DM detection~\cite{PandaX-4T:2021bab,LZ:2022lsv,XENON:2023cxc,PICO:2017tgi}. It turns out that if 
the lightest neutralino is bino-like and the bino mass parameter, $M_1$, is of the opposite sign as the $\mu$ parameter (related to the higgsino mass), the direct detection cross section is suppressed, opening the parameter space of possible LSP mass values~\cite{Ellis:2000ds,Ellis:2000jd,Ellis:2005mb,Baer:2006te,Huang:2014xua,Huang:2017kdh,Han:2018gej,Baum:2021qzx}. 

Moreover, sufficiently light supersymmetric particles could play a role in providing sizable additional contributions to the muon anomalous magnetic moment $g_\mu - 2$.  It has been shown in the co-annihilation scenario that demanding the wino mass parameter $M_2$ to be of the same sign as $\mu$ yields a positive contribution to  $g_\mu - 2$, thereby alleviating the apparent discrepancy between experimental results and theoretical calculations~\cite{Aoyama:2020ynm,Aoyama:2012wk,Aoyama:2019ryr,Czarnecki:2002nt,Gnendiger:2013pva,Davier:2017zfy,Keshavarzi:2018mgv,Colangelo:2018mtw,Hoferichter:2019mqg,Davier:2019can,Keshavarzi:2019abf,Kurz:2014wya,Melnikov:2003xd,Masjuan:2017tvw,Colangelo:2017fiz,Hoferichter:2018kwz,Gerardin:2019vio,Bijnens:2019ghy,Colangelo:2019uex,Blum:2019ugy,Colangelo:2014qya,Muong-2:2021ojo}.
Combining both conditions: $\mu \times M_1 < 0$ and $\mu \times M_2 > 0$, it happens that  the second lightest neutralino tends to decay radiatively into the lightest neutralino and a photon~\cite{Haber:1988px,Ambrosanio:1996gz,Baer:2002kv} in a significant way. Therefore, the searches for
radiative decays of neutralinos are quite intriguing and well motivated~\cite{Baum:2023inl}; and should be pursued at the LHC (A  similar enhanced radiative decay behavior has been pointed out in the  next-to-minimal-MSSM, see Ref.~\cite{Roy:2024yoh}).

Electroweakinos, as they are generically called the weakly interacting supersymmetric partners of the Higgs and electroweak gauge bosons,  are difficult  particles to detect at the LHC. This is specially the case when they decay radiatively and to a dark matter particle, due to the large Standard Model (SM) backgrounds. A naive analysis, based on simple cuts of the main kinematic variables, generally fails to show discovery potential even at the high luminosity LHC. These analyses, however, neglect the correlation between different kinematic variables, which are most accurately captured using machine-learning (ML) methods (for reviews of ML applied to high energy physics see, for instance,~\cite{Larkoski:2017jix,Mehta:2018dln,Guest:2018yhq,Albertsson:2018maf,Radovic:2018dip,Carleo:2019ptp,Bourilkov:2019yoi,Feickert:2021ajf,Schwartz:2021ftp,Karagiorgi:2021ngt,Coadou:2022nsh,Shanahan:2022ifi,Plehn:2022ftl,Belis:2023mqs,Bardhan:2024zla,Mondal:2024nsa,Halverson:2024hax}).

In this article, we perform and compare cut-based and ML analyses of the search for radiative neutralino decays at the LHC, for a center-of-mass energy of $\sqrt{s}$ = 14 TeV and a total integrated luminosity of ${\cal L} =$ 100 fb$^{-1}$. The analysis is a projection of the results that can be obtained in the early stages at the Run 3 of the LHC. Additionally, this choice for the luminosity saves computational resources in terms of Monte Carlo simulations.  For the ML analysis, we estimate discovery reach both within the traditional Binned-Likelihood (BL) method~\cite{Cowan:2010js} and within the Machine-Learned Likelihoods (MLL) method~\cite{Arganda:2022qzy,Arganda:2022mrd,Arganda:2022zbs,Arganda:2023qni}. In the latter, we perform an unbinned fit of the likelihood function obtained from the ML output using Kernel Density Estimators (KDE)~\cite{RosenblattKDE,ParzenKDE}.
We assume that all colored supersymmetric particles are heavy and beyond LHC reach,  and we focus on the production of winos, that in these MSSM scenarios are the second neutralino and the lightest chargino, and their radiative decay to the lightest neutralino and a photon. Due to the small mass difference between $\tilde \chi_1^0$ and $\tilde \chi_2^0/\tilde \chi_1^\pm$ required by co-annihilation, most of the signatures in the proposed signal process tend to be soft, so we consider the production of a highly energetic initial state radiation (ISR) jet in association with the electroweakino pair \cite{Baum:2023inl}. This increases the missing energy signature, which becomes useful in the search process. Contrary to the naive cut-and-count approach, the ML analysis shows an interesting discovery potential for electroweakinos at the LHC in the coming years. 

This article is organized as follows: in Section~\ref{sec:th-frame} we present the theoretical framework, highlighting the supersymmetric parameter space under consideration. The LSP is a bino-like neutralino that is almost degenerate in mass with  the wino-like neutralino $\tilde \chi_2^0$ and chargino $\tilde \chi_1^\pm$, and the radiative decay of $\tilde \chi_2^0$ is significant. Section~\ref{sec:collider} is dedicated to the collider analysis we conducted. First, we explain the event simulation and characterize the signal against the background. Then, we present the two approaches considered for the analysis: one based on rectangular cuts, optimized through a sequential cut selection, and a second approach using machine learning, considering both the BL method and the unbinned MLL method for the final significance estimation. The results of our collider analysis are shown in Section~\ref{sec:res}, comparing the potential of the different strategies applied. We reserve Section~\ref{sec:concl} for  our  conclusions. Appendix~\ref{app:complementaryplots} is devoted to complementary kinematic distributions of signal and background features used in our analysis. Appendix~\ref{app:blmll} reviews the main formulae for the BL and the MLL methods. Finally, in Appendix~\ref{app:trigger2}, we show that our conclusions are not significantly affected by considering a more stringent cut in the missing transverse energy, and we present an estimation of the impact of  systematic uncertainties in our results.

\section{Theoretical Framework}\label{sec:th-frame}
 
As mentioned in the introduction, several works have shown that there exist viable regions of the MSSM parameter space that simultaneously  reproduce the observed DM relic density of the universe and yield a solution to the apparent tension between theory and experiment in the value of the muon anomalous magnetic moment. Addressing both of these questions restricts  the MSSM  parameter space substantially. For example, a stable $\mathcal O(200)$\,GeV bino-dominated neutralino produced via freeze-out would be too abundant and severely violate DM relic density constraints. The right abundance is obtained if we consider the ``compressed region'' of parameter space, where there are two species that are almost degenerate in mass and the lightest of the two is a DM candidate. This allows for coannihilation to occur between these two species in the early Universe, reducing the LSP relic density to be in the ballpark of  the observed value.
At the same time, the sleptons play an important role in the MSSM contributions to $g_\mu - 2$, and therefore they cannot be too massive.
Resolution of the muon $g_\mu - 2$ apparent tension in the scenario of a bino-like LSP requires generically that the slepton masses $M_{\tilde l} \lesssim\, 1\; \rm{TeV}$~\footnote{Throughout this article, we shall assume the value of the muon $g_\mu - 2$ obtained by computing the hadronic vacuum polarization contributions by dispersion relations of the electron-positron hadronic cross sections~\cite{Borsanyi:2020mff}. If the final value tends to be the one obtained by lattice methods (see for instance Ref.~\cite{Borsanyi:2020mff} and Ref.~\cite{Coyle:2023nmi} for a discussion of this issue), the tension would be reduced and higher slepton masses than the ones assumed in this article would be preferred.}. 
This requirement on the slepton sector adds to the previously mentioned condition that $(M_2 \times \mu) >0$.
Finally, there exist direct detection limits that constrain neutralino relics scattering off nuclei via the exchange of a CP-even Higgs. 
This constraint is weakened for $(M_1 \times \mu) < 0$, where the spin-independent direct detection cross sections receive contributions from MSSM diagrams which partially cancel.

Motivated by the previous considerations, we are led to a region of parameter space that contains an interesting and yet unexploited, new collider signal. In this region, the LSP is a bino-like neutralino, $\tilde \chi_1^0$, which has a mass of several hundred GeV. Close in mass, but slightly heavier, are the next mass eigenstates of the electroweakinos, the second neutralino ($\tilde \chi_2^0$) and the lightest chargino ($\tilde \chi_1^\pm$). In addition, the sleptons must have masses of several hundred GeV, to fulfill current LHC bounds while still allowing for a possible contribution to $g_\mu - 2$.  Lastly, the higgsino masses need to be well below the TeV~scale, to avoid inducing a large bino-induced negative contribution to the muon $g_\mu - 2$~\cite{Baum:2021qzx}. For this study, we fix $M_{\tilde l} = 600\,$GeV and $|\mu| = 800$\,GeV. 

Typically, for the neutralino mass splitting larger than the observed Higgs mass, $m_H \sim 125$~GeV, $m_{\tilde \chi_2^0} - m_{\tilde \chi_1^0} \gtrsim m_H$,  the second neutralino decays via on-shell Higgs or electroweak gauge bosons. Decays of this type produce multilepton final states, which have been searched for extensively 
~\cite{ATLAS:2018ojr,ATLAS:2019lff,CMS:2020bfa,ATLAS:2021moa,CMS:2021cox,CMS:2021few,ATLAS:2021yqv,ATLAS:2022zwa,ATLAS:2019lng,ATLAS:2017vat,ATLAS:2019wgx}. However, the proper relic density can only be achieved deeper into the compressed region, for $m_{\tilde \chi_2^0} - m_{\tilde \chi_1^0} \sim 20-40$\,GeV. Interestingly enough, both collaborations have seen some multilepton excesses in potential electroweakino production  compatible with a compressed spectra and masses in the $200-300$~GeV range~\cite{ATLAS:2019lng, ATLAS:2021moa, CMS:2021edw}. There have been a few works trying to give a coherent explanation of these excesses~(see for example~\cite{Agin:2024yfs,Chakraborti:2024pdn}). In~\cite{Baum:2023inl} it was shown that in this region, due to the small mass differences between the second lightest and the lightest neutralino, the decays via off-shell electroweak gauge bosons become suppressed and give way to a radiative decay mode ($\tilde \chi_2^0 \to \tilde \chi_1^0 + \gamma$). Because this new mode is a 1-loop process with virtual sleptons and electroweakinos, it scales differently with the neutralino mass splitting and it is sensitive to several parameters in the MSSM parameter space, in particular the slepton masses. 
As mentioned before, the value of the slepton masses plays an important role in the radiative decay branching ratio of the second lightest neutralino, and larger slepton mass values will reduce it. Note that the value of $\tan(\beta)$ does not play a relevant role on the radiative decay branching ratio.~\footnote{The chosen value of slepton masses, $M_{\tilde l} = 600$\,GeV, and $\tan(\beta)$= 50 reduce the reported tension between the experimental measurement of $g_\mu - 2$ and theoretical predictions based on data-driven approaches. 
However, if one increases the slepton masses or lowers the value of $\tan(\beta)$ one could easily minimize the supersymmetric contributions to $g_\mu - 2$, thereby disregarding the current apparent tension in this quantity.}
 
The region of parameter space with significant, loop induced radiative decay offers a new signature for electroweakino searches at the LHC, which, as mentioned before, is dominated by wino production,  $pp \to \tilde\chi_2^0 \tilde\chi_1^\pm$. Current LHC searches that consider this production channel probe the compressed region by searching for multilepton final states. As proposed in~\cite{Baum:2023inl} it may prove beneficial to consider a soft photon ($p_T \sim \mathcal O(50)$\,GeV) as additional evidence for MSSM activity. In this paper we explore the experimental viability of searching for these photons  offering insights into the LHC reach potential   in the radiative decay mode. 

\begin{table}[t]
\centering
\begin{tabular}{c|c c c c c c c c }
  \toprule
  \textbf{BP} & \makecell{$m_{\tilde \chi_2^0}$ \\ {[GeV]}} & \makecell{ $(m_{\tilde \chi_2^0} - m_{\tilde \chi_1^0})$\\ {[GeV]}} & Br$(\tilde \chi_2^0 \to \tilde\chi_1^0\gamma)$ & $\sigma(p p \rightarrow \tilde \chi_1^\pm \tilde \chi_2^0)$ &  $\frac{\sigma_{DD} \times \min(\Omega_{\tilde \chi_1^0}, \Omega_{\rm DM})}{\sigma_{DD,95} \times \Omega_{\rm DM}}$ & $\Omega_{\tilde \chi_1^0} h^2$ & \makecell{$\Delta a_\mu^{\rm MSSM}$ \\ $\times 10^{10}$} \\
  \midrule
  \addlinespace
    \bf 1 & 200 & 34 & 15\% & 190 fb & 0.34 & 2.0  & 22 \\
    \rowcolor{Gray!20}
    \bf 2 & 200 & 19 & 37\% & 190 fb & 0.47 & 0.12  & 22 \\
    \bf 3 & 200 & 10 & 73\% & 190 fb & 0.09 & 0.02  & 22 \\
    \bf 4 & 250 & 37 & 15\% & 92 fb & 0.74 & 1.2 & 21 \\
    \rowcolor{Gray!20}
    \bf 5 & 250 & 22 & 36\% & 92 fb & 0.96 & 0.12 & 21 \\
    \bf 6 & 250 & 13 & 67\% & 92 fb & 0.19 & 0.03 & 21 \\
    \bf 7 & 300 & 39 & 16\% & 48 fb & 1.3 & 0.8 & 20 \\
    \rowcolor{Gray!20}
    \bf 8 & 300 & 24 & 36\% & 48 fb & 1.6 & 0.12 & 20 \\
    \bf 9 & 300 & 15 & 62\% & 48 fb & 0.36 & 0.04 & 20 \\
    \bf 10 & 350 & 41 & 17\% & 27 fb & 2.2 & 0.60 & 19 \\
    \rowcolor{Gray!20}
    \bf 11 & 350 & 26 & 35\% & 27 fb & 2.5 & 0.12 & 19 \\
    \bf 12 & 350 & 17 & 58\% & 27 fb & 1.0 & 0.04 & 19\\
    \bf 13 & 400 & 43 & 16\% & 16 fb & 3.2 & 0.47 & 17 \\
    \rowcolor{Gray!20}
    \bf 14 & 400 & 27 & 32\% & 16 fb & 3.6 & 0.12 & 17\\
    \bf 15 & 400 & 18 & 52\% & 16 fb & 1.7 & 0.05 & 17 \\
  \bottomrule
\end{tabular}
\caption{Table of benchmark points used in this analysis. For all points, $\mu = 800$\,GeV, $\tan(\beta) = 50$, and $M_{\tilde l} = 600$\,GeV, and $(M_1 \times \mu) < 0$. We set the squark soft masses to 2.5\,TeV, and any additional superpartner masses to 1.5\,TeV. Also, $m_{\tilde \chi_1^\pm} \approx m_{\tilde \chi_2^0}$ for all of the points. The points which produce the proper cosmological relic density are shaded in gray. Direct detection cross sections for points which produce relic $\tilde \chi_1^0$ abundance above $\Omega_{\rm DM}$ are rescaled to estimate the local abundance of $\tilde \chi_1^0$. The preferred value of $\Delta a_\mu$ is $25 \pm 5$. A few of the low-mass points are excluded by ATLAS of CMS \cite{CMS:2024gyw,ATLAS:2019lng, ATLAS:2021moa, CMS:2021edw}. Points are chosen in a grid across the soft parameters $M_2$ and $M_2 - M_1$ which leads to a slightly uneven set of splittings as there is a variation of the dependence of physical parameters on soft parameters.}
\label{table:bps}
\end{table}
 
We consider 15 benchmark points (BPs) whose relevant parameters for collider phenomenology are shown in Table~\ref{table:bps}. We generate the mass spectra and decay rates for these points using \verb|SuSpect|~\cite{Kneur:2022vwt} and \verb|SUSY-HIT|~\cite{Djouadi:2006bz}. Cross sections were generated using \verb|PROSPINO|~\cite{Beenakker:1996ed}, and the direct detection constraints and relic density were evaluated with \verb|MicrOmegas|~\cite{Belanger:2018ccd}, including the recently published LUX-ZEPLIN results~\cite{LZCollaboration:2024lux}.
These benchmark points are chosen as representative points across the compressed region. The relic density is directly dependent on the mass splitting, and so not all the BPs necessarily produce the cosmological relic density, but a subset does (BPs 2, 5, 8, 11, and 14). Points predicting a relic density above the observed one are excluded within the MSSM but may be brought back into agreement by additional SM modifications, such as a late injection of entropy during the Big Bang or another species with which $\tilde \chi_1^0$ can coannihilate. For those points, we rescale the direct detection cross section by the predicted relic density to account for a modified local abundance, consistent with the observed one.   A few of these BPs, with LSP below 250~GeV, are in principle excluded by previous searches. In particular, the benchmarks with lowest neutralinos masses would be excluded by ATLAS and CMS multilepton searches (e.g. \cite{ATLAS:2019lng, ATLAS:2021moa, CMS:2021edw}).  However, the ATLAS and CMS bounds applied to our scenario are weakened by the change in the assumptions. For example, CMS assumes both 100\% branching ratios of $(\tilde \chi_2^0 \to \tilde \chi_1^0 + h^*)$ and higher squark masses, which are modified significantly in our BPs. This affects both the rates of the multilepton final states and the production cross section for $(pp\to\tilde \chi_2^0\tilde \chi_1^\pm)$. For the LHC bounds depicted in Figure~\ref{fig:contour_plots}, we use the bounds placed in~\cite{Chakraborti:2024pdn}, which provide a more recent update to the bounds in~\cite{Baum:2023inl} for the same region in parameter space.

Due to the large non-standard Higgs boson masses allowed by LHC searches for large values of $\tan\beta$~\cite{CMS:2018rmh,ATLAS:2020zms}, there is only a small suppression of the direct DM detection rate induced by the heavy Higgs bosons exchange and the direct DM detection amplitude is dominated by the  SM-like CP-even Higgs boson exchange. A larger suppression may be obtained for smaller $\tan \beta$ values ($\tan\beta  \leq {\cal{O}}(10)$), for which the LHC bounds become weaker and the blind spot conditions~\cite{Huang:2014xua} 
\begin{equation}
\frac{2(M_1 + \mu \sin2\beta)}{m_h^2} \simeq -\frac{\mu \tan\beta}{m_H^2}
\end{equation}
may be approached by lowering the heavy CP-even Higgs boson mass $m_H$. The supersymmetric $g_\mu - 2$ contribution becomes subdominant in this region of mass parameters. A dedicated analysis of the allowed region of parameters would be interesting, but it is beyond the scope of this article.

\section{Collider Analysis}\label{sec:collider}

\subsection{Event Simulation and Characterization}
\label{sec:events}

\begin{figure}
    \centering
    \includegraphics[width=0.6\textwidth]{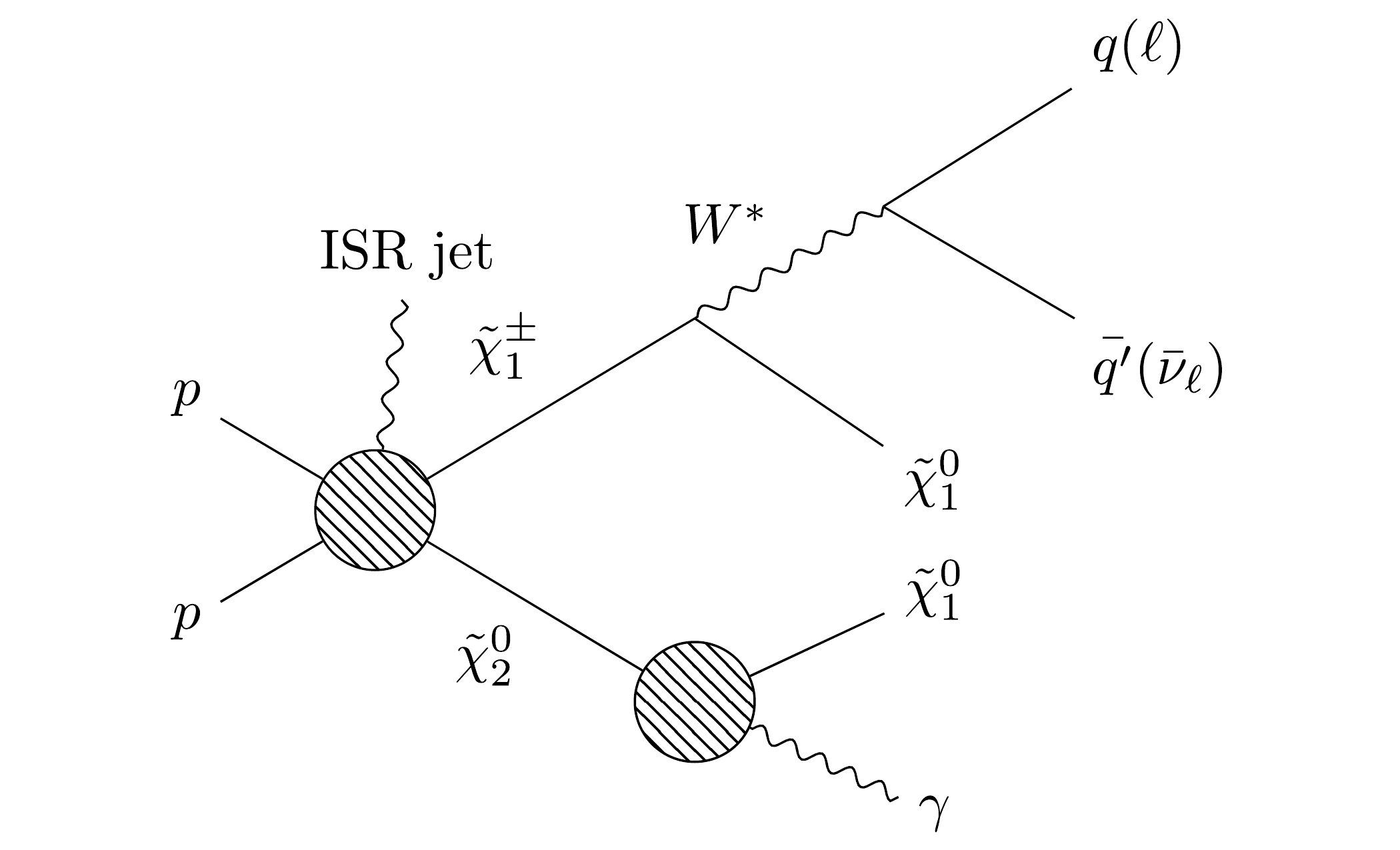}
    \caption{Feynman diagram of the production mode and final state considered.}
    \label{fig:feynman-diagram}
\end{figure}

The proposed signal is the LHC production of the lightest chargino, $\tilde \chi_1^{\pm}$, and the second-lightest neutralino, $\tilde \chi_2^0$, (both wino-like) in addition to an initial state radiation jet, as can be seen in Figure~\ref{fig:feynman-diagram}. We consider that the chargino decays to the lightest neutralino plus a lepton and a neutrino through an off-shell $W$ boson, $ \tilde \chi_1^{\pm} \to \tilde \chi_1^0 \ell \nu_{\ell}$, while the second lightest neutralino decays radiatively to the lightest neutralino plus a photon, $\tilde \chi_2^0 \to \tilde \chi_1^0 + \gamma$. Therefore the LHC signal process under study is the following:
\begin{equation}
pp \to \tilde \chi_1^{\pm} \, \tilde \chi_2^0 \, j \to \tilde\chi_1^0 \, \ell \, \nu_{\ell} + \tilde\chi_1^0 \, \gamma + j \, .
\end{equation}

Signal events have been generated with {\tt MadGraph5\_aMC@NLO}~\cite{Alwall:2014hca}, using the Universal Feynman Output (UFO)~\cite{Darme:2023jdn} model {\tt MSSM\_SLHA2}~\cite{Duhr:2011se} at LO in QCD with the NNPDF2.3 LO PDF set \cite{Ball:2012cx}, for an LHC center-of-mass energy of 14 TeV. Parton showering and hadronization processes have been performed with {\tt Pythia8}~\cite{Sjostrand:2014zea,Sjostrand:2007gs}, while {\tt Delphes}~\cite{deFavereau:2013fsa} has been used for a fast detector simulation response, using the default ATLAS card~\footnote{For the sake of simplicity, we focus our analysis in the ATLAS detector, but results are expected to be extensive to CMS experiment.}. Spin correlations in the decays of the charginos $\tilde \chi_1^\pm \to \tilde \chi_1^0 \, \ell^\pm \,\nu_{\ell}$ ($\ell=e, \mu$) have been taken into account via the {\tt MadGraph5\_aMC@NLO} decay chain syntax.
We have simulated events within the mass range $m_{\tilde\chi_2^{0}}\in [200,400]$ GeV, selecting BPs with a step of $50$ GeV in $m_{\tilde\chi_2^{0}}$ as shown in Table~\ref{table:bps}. For each value of $m_{\tilde\chi_2^{0}}$, we explore three different values of $m_{\tilde\chi_2^{0}}-m_{\tilde\chi_1^{0}}$.

Background events have been simulated at LO in QCD with {\tt MadGraph5\_aMC@NLO}, and subsequently have been processed with {\tt Pythia8} and {\tt Delphes}. We have considered the contributions to the mono-photon search channel, characterized by the presence of large missing transverse energy, a highly boosted initial state radiation jet, and the presence of at least one charge lepton. The dominant channels are $W+$jets, $W\gamma$ and $t\bar{t}+$jets. Additionally, we also considered in our sample other subdominant contributions, including $Z+$jets, single-top, $t\Bar{t}\gamma$ and diboson ($ZZ$, $WW$, and $ZW$). Other contributions are expected to be negligible. 

Regarding object identification criteria for light leptons, electron candidates are required to have $p_T>10$ GeV and $|\eta|<2.47$, and should also lie outside the transition region ($1.37<|\eta|<1.52$) between the barrel and endcap calorimeters of the ATLAS detector, while muons must have $p_T>10$ GeV and $|\eta|<2.7$. Jets are required to satisfy $p_T>20$ GeV and $|\eta|<4.5$. Hadronically decaying $\tau$ leptons are required to have $p_T>20$ GeV and $|\eta|<2.47$, excluding the $\eta$ range $1.37<|\eta|<1.52$. 
Photon candidates are required to have $p_T>10$ GeV and $|\eta|<2.37$. 

Event selection criteria require at least one charged light lepton ($\ell=e, \mu$), at least one photon, and at least one jet. The leading jet, typically the ISR jet, is required to have $p_T>100$ GeV. Due to the presence of neutralinos and neutrinos in the final state, we require $E_T^\text{miss}>100$ GeV. We have checked that requiring $E_T^\text{miss}>200$ GeV do not change significantly our results (see Appendix~\ref{app:trigger2}). In this work, we propose a looser trigger for the HL-LHC which implies a larger number of expected events and could allow a better determination of uncertainties, that could be implemented by the experimental LHC collaborations in the searches for this class of DM signals.

The expected signal and background events yields estimated for the LHC at a center-of-mass energy of 14 TeV and a total integrated luminosity of 100 fb$^{-1}$ after applying event selection criteria can be found in Table~\ref{tab:numberofevents}. For all the BPs selected in the MSSM parameter space, the expected yield is $\sim 3$ orders of magnitude lower than the expected total background yield. A first approximation of the discovery significance, $S/\sqrt{B}$, in each case, is shown in the third column of Table~\ref{tab:numberofevents}, where $S$ and $B$ are the expected numbers of signal and background events, respectively. Neither evidence nor discovery is achieved in any of the examples. 

\begin{table}[t]
    \centering
    \begin{tabular}{c|c c c|c|c}
       \cline{1-2} \cline{4-6}
       Process  & Yield &  \hspace{1cm} & BP \# & Yield & $S/\sqrt{B}$ \\
       \cline{1-2} \cline{4-6}
        $W+\text{jets}$        & $ 60058$ & & 1 & $202$ & $0.52$\\
        $W\gamma$              & $ 58462$ & & 2 & $459$ & $1.19$\\
        $t\Bar{t}+\text{jets}$ & $ 18051$ & & 3 & $637$ & $1.65$\\
        $Z+\text{jets}$        & $ 3360$  & & 4 & $111$ & $0.28$\\
        $\text{Single-top}$             & $ 3214$  & & 5 & $235$ & $0.61$\\
        $t\Bar{t}\gamma$       & $ 2498$  & & 6 & $334$ & $0.86$\\
        $\text{Diboson}$                   & $ 2340$  & & 7 & $66$  & $0.17$\\
   
        \cline{1-2}
        Total background       & $ 147983$& & 8 & $129$ & $0.33$\\
        \cline{1-2}
        \multicolumn{2}{c}{\multirow{8}{1cm}{}} & & 9 & $ 179$ & $0.46$ \\
        \multicolumn{2}{c}{\multirow{8}{1cm}{}} & & 10 & $ 40$ & $0.10$ \\
        \multicolumn{2}{c}{\multirow{8}{1cm}{}} & & 11 & $ 74$ & $0.19$ \\
        \multicolumn{2}{c}{\multirow{8}{1cm}{}} & & 12 & $102$ & $0.26$ \\
        \multicolumn{2}{c}{\multirow{8}{1cm}{}} & & 13 & $ 23$ & $0.05$ \\
        \multicolumn{2}{c}{\multirow{8}{1cm}{}} & & 14 & $ 41$ & $0.10$ \\
        \multicolumn{2}{c}{\multirow{8}{1cm}{}} & & 15 & $ 57$ & $0.14$ \\
        \cline{4-6}
    \end{tabular}
    \caption{Expected background and signal events at the LHC with a center-of-mass energy of 14 TeV and a total integrated luminosity of  100 fb$^{-1}$, after applying event selection criteria.}
    \label{tab:numberofevents}
\end{table}

After applying the baseline event selection criteria, we have explored a large but simple set of variables to characterize the kinematics of the studied final state, and subsequently to design our cut-based and machine-learning strategies. We considered not only fundamental low-level detector variables such as the transverse momentum ($p_T$) and pseudorapidity ($\eta$) of the leading jet ($j_1$), lepton ($\ell_1$), and photon ($\gamma_1$), the missing transverse energy ($E_T^\text{miss}$) and object multiplicities, but several high-level observables that contribute to a more nuanced understanding of the underlying physics and play a crucial role enhancing the signal-to-background discrimination.
Among these observables we incorporate the hadronic activity, defined as 
\begin{equation}
    H_T^{\text{jets}}=\sum p_T^{\text{jets}} \,,
\end{equation}
which provides information about
the formation of hadronic particles in the final state; the total transverse energy, a measure of the hardness of the event~\cite{ATLAS:2018uid}, defined as
\begin{equation}
    H_T=\sum p_T^{\text{jets}} + \sum p_T^{\tau} + \sum p_T^{e} + \sum p_T^{\mu} + \sum p_T^{\gamma} \, ;
\end{equation}
the transverse mass of each leading particle $A=\{j_1, \ell_1, \gamma_1 \}$, 
\begin{equation}
    m_T^{A}\equiv m_T\left(\mathbf{p}_T (A), \mathbf{E}_T^\text{miss}\right)=\sqrt{ 2 p_T(A) E_T^\text{miss} \left(1-\cos \Delta \phi \left(\mathbf{p}_T (A), \mathbf{E}_T^\text{miss}\right)\right)} \,,
\end{equation}
where $p_T(A) = |\mathbf{p}_T (A)|$, and $E_T^\text{miss} = |\mathbf{E}_T^\text{miss}|$; the scalar sum of the transverse momentum of all the leading particles,
\begin{equation}
    s_{T}^{1}=p_T^{\ell_{1}}+p_T^{j_{1}}+p_T^{\gamma_{1}} \, ;
\end{equation}
and the missing transverse energy significance, defined as $E_T^\text{miss}/\sqrt{H_T}$. A high value of the latter variable indicates that the observed $E_T^{\text{miss}}$ cannot be explained by momentum resolution effects, then the event is more likely to contain undetected objects such as neutrinos or more exotic invisible particles~\cite{collaboration_2011}. 

\begin{figure}
    \centering
    \includegraphics[width=1\textwidth]{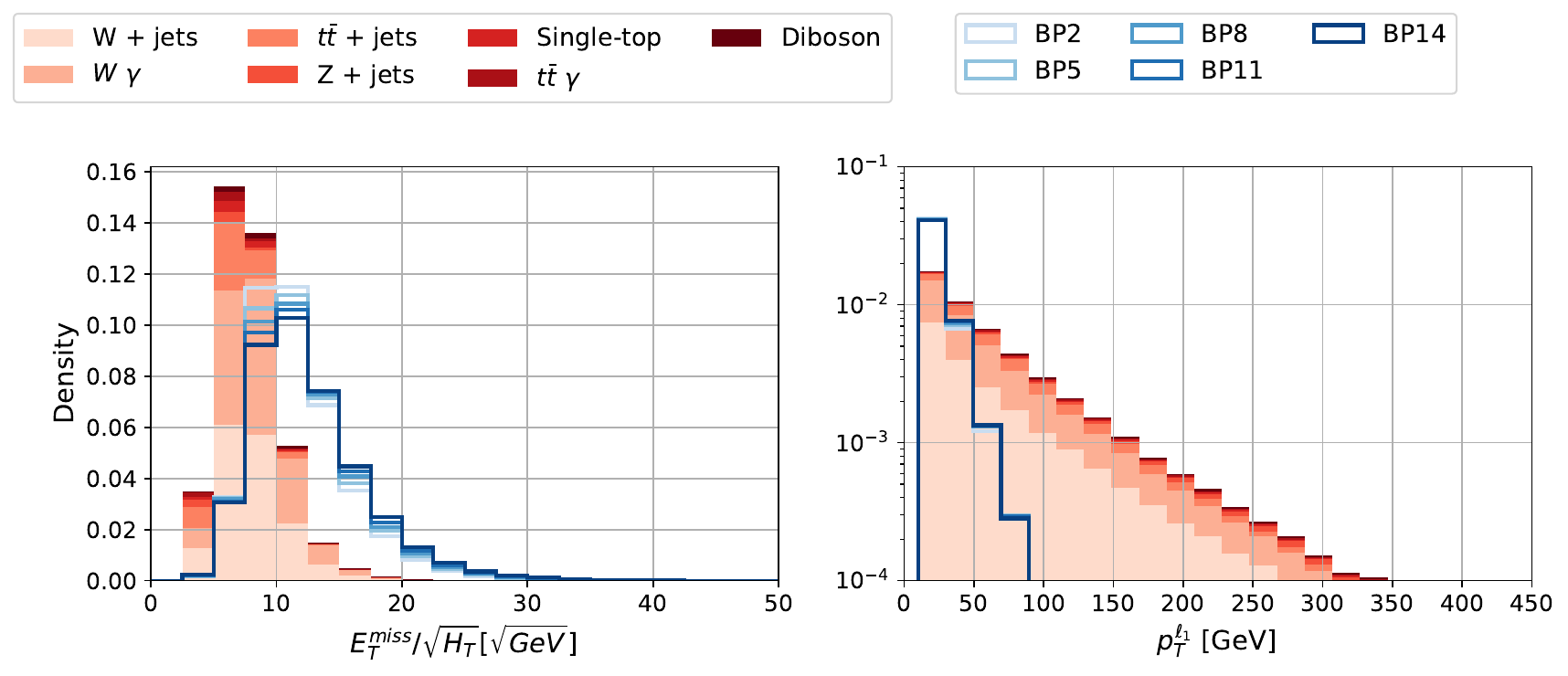}
    \includegraphics[width=1\textwidth]{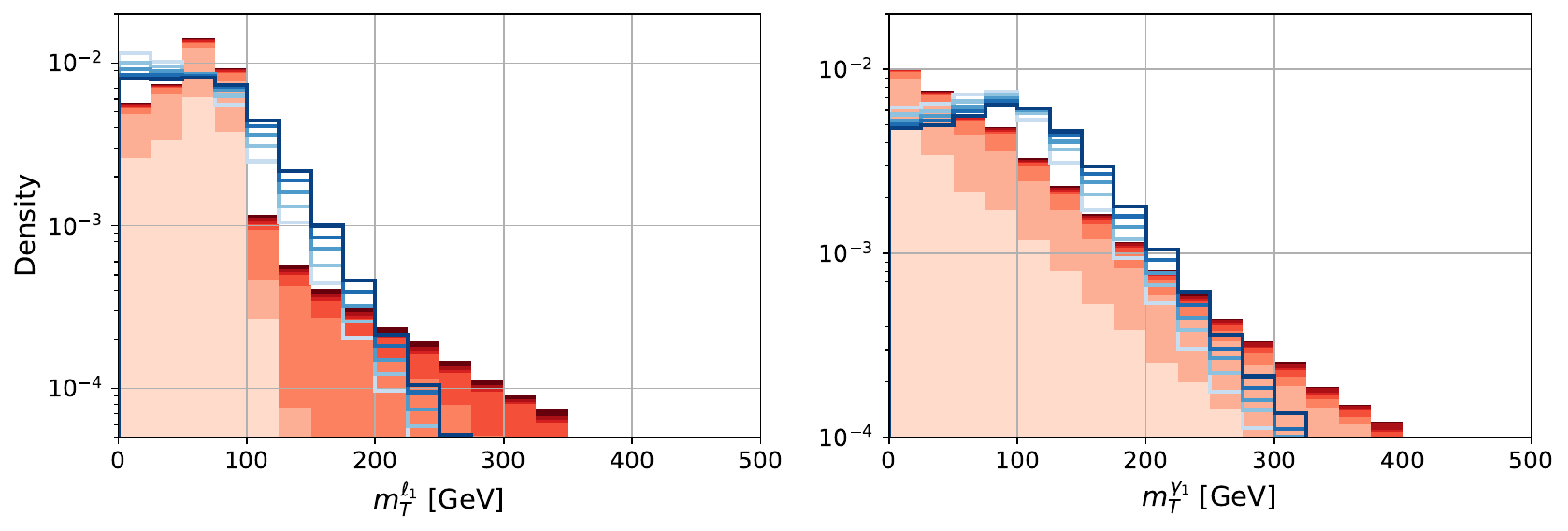}
    \caption{Distributions of the four final state kinematic variables showing the most discrimination between signal and background: missing transverse energy significance, $E_T^\text{miss}/\sqrt{H_T}$ (top-left panel); total transverse momentum of the leading lepton, $p_T^{\ell_1}$ (top-right panel); transverse mass of the leading lepton, $m_T^{\ell_1}$ (bottom-left panel); and transverse mass of the leading photon, $m_T^{\gamma_1}$ (bottom-right panel). We display a subset of the benchmark points, choosing those which produce the correct relic density.}
    \label{fig:relevant-variables}
\end{figure}

In Figure~\ref{fig:relevant-variables} we present the distributions of the four most relevant variables for the discrimination task performed in the next section. The total SM background incorporates all the processes in Table~\ref{tab:numberofevents}, considering the weight of each channel by their relative cross sections. As an example, we show 
5 different BPs that saturate the measured amount of DM relic density, while the rest of the BPs have similar behavior. 
We have found that the most discriminant variable is the missing transverse energy significance (shown in the top-left panel), partially inherited from the harder $E_T^{\text{miss}}$ distribution of the signal events given by the recoil of the $(\Tilde{\chi}_{2}^{0}+\Tilde{\chi}_{1}^{\pm})$ system against the ISR jet. Since $m_{\tilde\chi_1^{0}} \simeq m_{\tilde\chi_2^{0}} \simeq m_{\tilde\chi_1^{\pm}} \sim \mathcal{O}(200-400\text{GeV})$, we expect $E_T^{\text{miss}}$ dominated by the contribution of the two lightest neutralinos in the final state. We obtain a high value of $E_T^\text{miss}/\sqrt{H_T}$ both for background and signal, manifesting the presence of undetected objects. In the case of the SM background, the only physical contribution comes from neutrinos produced in $W$ decays which escape the detector, while for the signal there is a large contribution from the heavy invisible neutralinos.

The second most important feature, and the main low-level variable, 
is the transverse momentum of the leading lepton ($p_T^{\ell_{1}}$), shown in the top-right panel of Figure~\ref{fig:relevant-variables}. 
Since the masses of the lightest chargino $\tilde \chi_1^\pm$ and the lightest neutralino $\tilde \chi_1^0$ are close, the leptons produced in the chargino decay through an off-shell $W$ gauge boson, $\tilde \chi_1^\pm \to \tilde\chi_1^0 \, \ell \, \nu_{\ell}$, have a soft distribution.
On the other hand, the leptons arise mostly from on-shell $W$-boson decays in the background case, with larger phase space. 

The transverse mass of the leading lepton (bottom-left panel of Figure~\ref{fig:relevant-variables}) is also important. 
In the considered SM processes, this variable peaks around the $W$-boson mass since the only contribution to $E_T^{\text{miss}}$ comes from the neutrinos.
The back-to-back configuration between the leading lepton and the reconstructed $E_T^{\text{miss}}$ is mostly favored for background events, where the neutrinos can recoil with the corresponding charged leptons from the SM $W$-boson decays. As mentioned above, the $E_T^{\text{miss}}$ in the signal process is dominated by the neutralinos, producing a broader $m^{\ell_1}_T$ distribution. 

Finally, on the bottom-right panel of Figure~\ref{fig:relevant-variables} we show the fourth most relevant feature, the transverse mass of the leading photon. In the signal case, this feature peaks around $100$ GeV, the minimum $p_T$ value set in the event selection criteria for the leading jet. Since the $(\Tilde{\chi}_{2}^{0}+\Tilde{\chi}_{1}^{\pm})$ system decays to missing energy plus a photon and a soft lepton (the latter can be seen in the top-right panel of Figure~\ref{fig:relevant-variables}), the transverse mass of the photon seems to reconstruct the system transverse energy that recoils against the ISR jet. On the other hand, in the background processes the photon transverse mass is not directly associated with the leading jet. 
The distributions for the rest of the low-level and high-level variables can be found in Figures~\ref{fig:low-level-app} and~\ref{fig:high-level-app} of Appendix~\ref{app:complementaryplots}.

\subsection{Analysis Strategies}

In this section, we present two different approaches for our LHC exclusion and discovery reach analyses: a sequential cut-based analysis, that optimizes cuts in all the variables considered in the signal and background characterization; and a machine-learning one, which, after using a binary supervised classifier for signal and background discrimination, computes the final discovery significance using the entire ML output by means of the BL method and the unbinned MLL method. It is important to note that each analysis is trained with events from the complete set of BPs, and hence, can be applied to all the compressed parameter space.

\subsubsection{Cut-Based Strategy}

To refine the cut-and-count method, we propose a sequential cut-based (SCB) strategy, that automatically searches the best cuts in all variables by assessing the impact of each cut to the overall significance.
This implies the exploration of a high dimensional space that can be computationally expensive.
In our case we have $22$ variables, and we need to define lower and upper cuts, leading to $44$ degrees of freedom. To simplify the problem, in this work we explore one variable at a time in an iterative way. Although this solution could be improved, this is beyond the scope of the paper as, in any case, the machine-learning approach described in the following section is more flexible than the rectangular cut-based method.

The SCB procedure is based on the following steps:
\begin{enumerate}
    \item Perform a grid with $10$ points evenly distributed in the available range, including its borders, of a randomly selected variable.
    \item Use each point as an upper cut for the signal region and estimate the significance, $Z = S/\sqrt{B}$, where $S$ and $B$ are the numbers of signal and background expected events below the threshold associated with that point.
    \item Use each point as a lower cut for the signal region and estimate the significance, $Z = S/\sqrt{B}$, where $S$ and $B$ are the numbers of signal and background expected events above the threshold associated with that point.
    \item Apply and save the cut (superior or inferior) that leaves the best significance and has more than $5$ signal and $5$ background events.
    \item Repeat items $1$ to $4$ taking into account the new updated range of each variable.
\end{enumerate}
We have performed $100$ iterations to allow each variable to be analyzed more than one time and hence the cuts can be updated. To make our analysis applicable to the entire parameter space, in steps $2$ and $3$ we have applied the upper and lower cut to all BP datasets and computed the resulting significance in each one. Then, the optimal cut decided in step $4$ takes into account the significance average over all the BPs. Although the final signal region would not be the optimal one for each BP, the approach is more suited to explore the parameter space of interest. 

\subsubsection{Machine-Learning Strategy}

For the ML analysis, we used a supervised per-event binary classifier based on the \texttt{XGBoost}~\cite{Chen:2016btl} toolkit, trained to distinguish between signal and background events
using as input features all the low-level and high-level variables described in Section~\ref{sec:events}. \texttt{XGBoost} or extreme gradient boosting is a ML method used in supervised problems either for classification or regression. It is also very popular in the high-energy physics community and widely used by ATLAS and CMS Collaborations. The algorithm is based on parallel tree boosting, i.e. it gives a prediction model in the form of an ensemble of weak prediction models (decision trees). 

For the background dataset, we employed $200$k events considering the relative weight of each background channel, while for the signal we used $200$k events (equal number of events per BP). In this way, we ensure to have balanced datasets for the training of the machine-learning models. Notice that we train a single ML model using samples from all BPs. While this approach may not be optimal for each particular BP, it is intentional. As mentioned before, the goal is to apply the same ML classifier across the entire parameter space, as the true underlying model would be unknown in an experimental search.
In order to avoid overfitting, we have split each dataset into two independent datasets: a training set consisting of $80\%$ of the original dataset; and a validation set consisting of the remaining $20\%$ data.
In addition, we have also prepared a third independent dataset, dubbed test set with $\sim400$ background and signal events. It is used to assess the performance of the algorithm once it is already trained and to compute the significances as explained below. 

\begin{figure}
  \centering
  \includegraphics[width=0.49\textwidth]{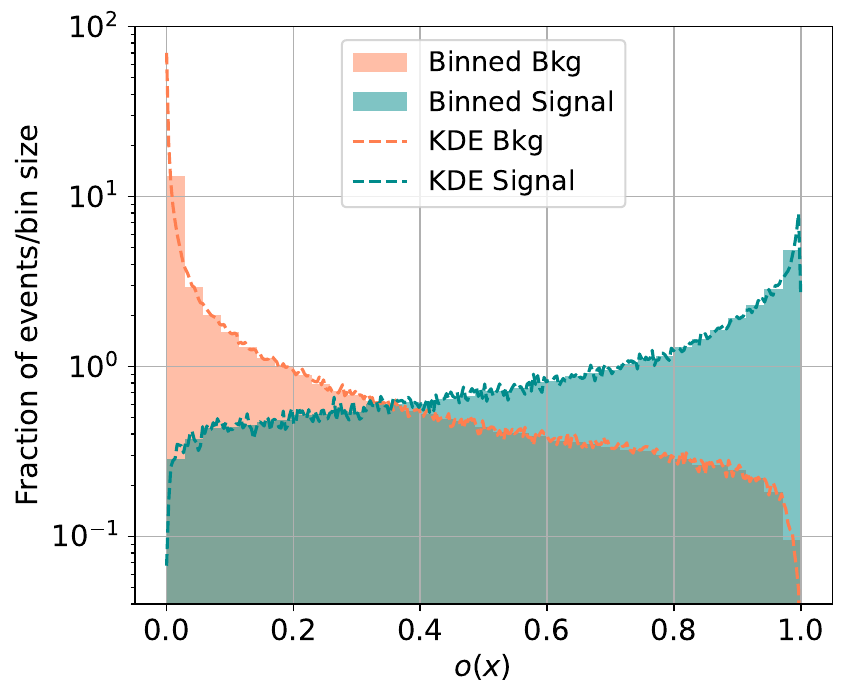}
  \includegraphics[width=0.49\textwidth]{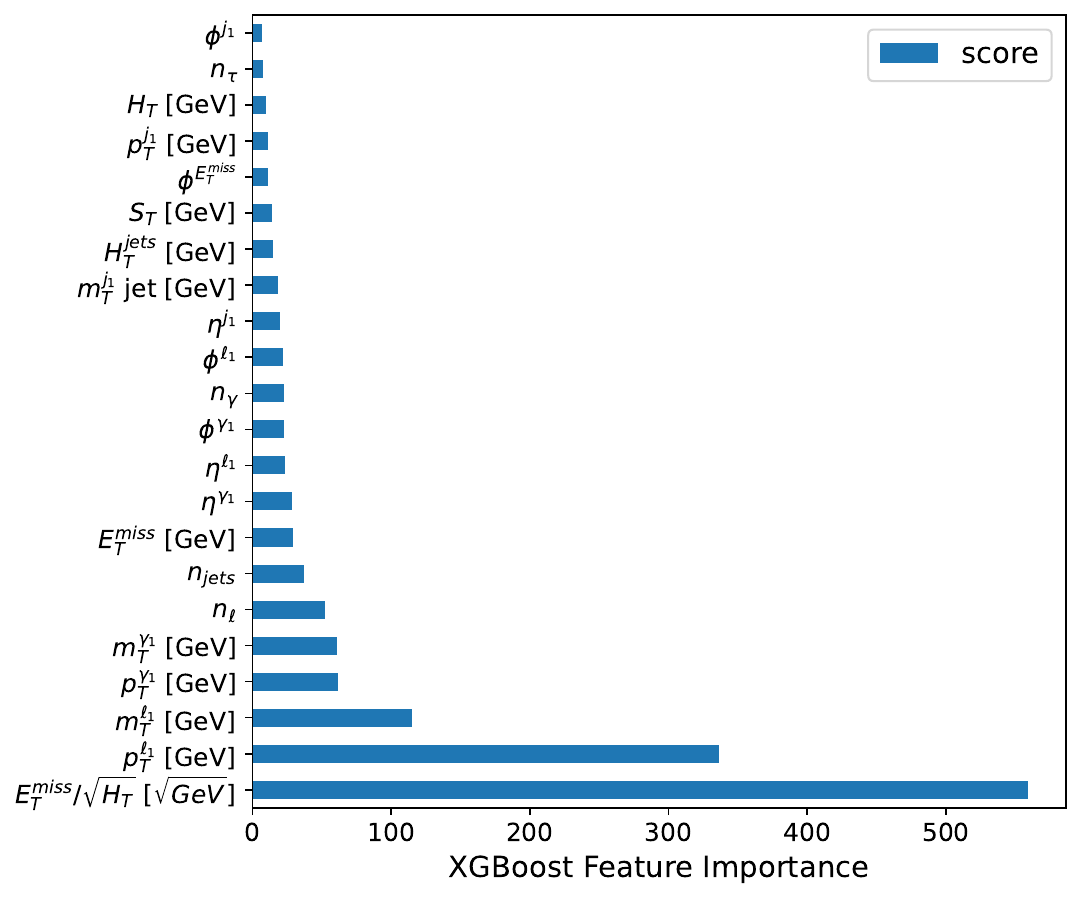}
\caption{ Left panel: output of the \texttt{XGBoost} classifier trained using all the BPs as signal (AUC$=0.94$), when tested with only pure background (orange) or pure signal (green) samples. The dashed curves correspond to the PDFs obtained with the KDE fit. Right panel:  feature importance score (gain metric) for the same \texttt{XGBoost} classifier.}
\label{XGBoost-outputs}
\end{figure}

The one-dimensional output of the binary classifier, $o(x) \in [0,1]$, quantifies the probability of being signal-like of a given event, i.e., for signal-like events, the output $o(x)$ should be close to 1 while for background-like events $o(x)$ should be close to 0. The output of the ML classifier can be seen in the left panel of Figure~\ref{XGBoost-outputs} when tested with only pure background or pure signal samples, orange and green histograms, respectively.
As expected, it can be seen that signal and background events peak at probabilities of $1$ and $0$ respectively, highlighting the classifier performance on the discrimination task (AUC$=0.94$).
For completeness, using the mentioned ML classifier, we analyzed its performance distinguishing the signal against each one of the backgrounds. We found that the discrimination powers are very similar with a slightly better performance considering the more abundant ones.
This comes from the fact that the training set was build weighting the number of events by the cross sections of each background process, hence the algorithm focuses more on those channels.
In the right panel of Figure~\ref{XGBoost-outputs} we show the feature importance score for the \texttt{XGBoost} classifier.
This magnitude is estimated from the gain metric that measures the relative contribution of the corresponding feature for generating a prediction. A higher value of this metric implies a higher impact on discrimination. The most relevant features, as already anticipated, are the missing transverse energy significance, $E_T^\text{miss}/\sqrt{H_T}$; the momentum of the leading lepton, $p_T^{\ell_{1}}$; the transverse mass of the leading lepton, $m_T^{\ell_1}$; and the transverse mass of the leading photon, $m_T^{\gamma_1}$. 

After the binary classification, the computation of the exclusion and discovery reach 
($Z$ = 2 or $Z$ = 5, respectively; with $Z$ defined in Appendix~\ref{app:blmll}) is done using the entire one-dimensional classifier output in both binned and unbinned approaches, based on the traditional BL method and the MLL method, respectively. The binning process inherently sacrifices information on probability densities within each bin, impacting the likelihood estimation and potentially reducing the significance. While reducing bin width can mitigate this loss, finite statistics render this approximation unreliable. In contrast, unbinned methods maintain individual data point granularity, offering a potentially more accurate data representation without averaging information across bins, although it may suffer from numerical fluctuations that in binned methods are smoothed out.

On one hand, for the BL approach, the likelihood function is built as the product of the Poisson probability functions modeling the population of each bin, as shown in Appendix~\ref{app:blmll}. Therefore the full ML output is binned and turned into a histogram to find the expected number of signal and background events, to finally compute the significance with a likelihood ratio test.

On the other hand, the unbinned approach applied in this work is based on the MLL framework~\cite{Arganda:2022qzy,Arganda:2022mrd,Arganda:2022zbs,Arganda:2023qni}. It uses KDE, a non-parametric method for density estimation (thus it does not assume a specific functional form for the underlying distribution), to fit the classifier output when tested with only pure background or pure signal samples, respectively. 
These fitted distributions are used as a one-dimensional approximation of the background and signal probability density functions (PDF), $\tilde{p}_{s}(o(x))$ and $\tilde{p}_{b}(o(x))$, and then introduced into an unbinned likelihood function to compute the significance with a likelihood ratio test, as can be seen in Appendix~\ref{app:blmll}.
The KDE fit only involves the bandwidth parameter, which controls the degree of smoothness of the estimated density function. To obtain this parameter, we have trained independent signal and background KDEs with $50$k input points each~\footnote{The density estimation tends to converge to the true underlying distribution for large datasets. We have checked that increasing the number of events used in this work for the KDE fits yields no significant changes.}. Then, the value of the bandwidth in each case has been found through a grid search employing the \texttt{GridSearchCV} function in the \texttt{sklearn.model\textunderscore selection} Python package~\cite{scikit-learn}. This method selects the bandwidth that maximizes data likelihood in a 5-fold cross-validation strategy. For further details, we refer the reader to the MLL references~\cite{Arganda:2022qzy,Arganda:2022mrd,Arganda:2022zbs,Arganda:2023qni}.

In the left panel of Figure~\ref{XGBoost-outputs} we also show, as dashed lines, the corresponding KDE fit for the background (orange lines) and signal (green lines) probability distributions.
As expected, it can be seen that both KDE fits describes the probability distributions of the signal and background ML outputs. These probabilities are used in turn to assess the significance of the MLL method as explained in Appendix~\ref{app:blmll}.

\section{Results}
\label{sec:res}

\begin{figure}
    \centering
    \includegraphics[width = 1\textwidth]{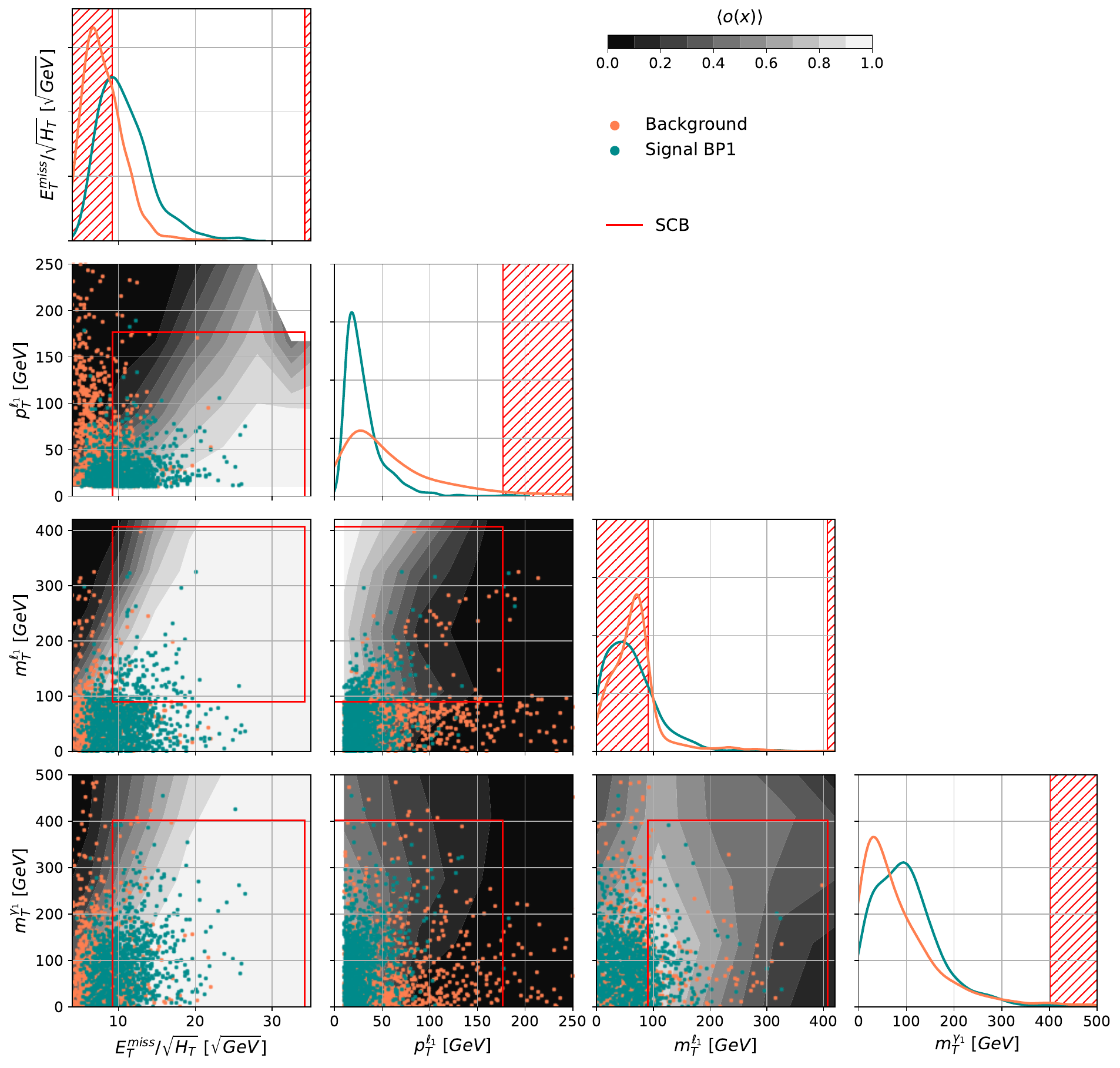}
    \caption{ Correlation plot of the four most important features according to the ML algorithm (see right panel of Figure~\ref{XGBoost-outputs}) for signal (green dots) and background (orange dots) events. 
    The gray scale corresponds to the average machine-learning output $\langle o(x) \rangle$ in 2D bins.
    The signal-enriched regions found with the SCB method are demarcated with solid red lines.
    In the diagonal we show the PDF of each feature distinguishing signal and background events. Finally, the hatched regions correspond to the excluded areas of the SCB method.}
    \label{fig:ImpVars_correlation}
\end{figure}

To compare the two analysis strategies, in Figure~\ref{fig:ImpVars_correlation} we present the SCB signal enriched regions and the ML classifier behavior, applied to the BP1 dataset as an example. To ease visualization we show only four features: $E_T^\text{miss}/\sqrt{H_T}$, $p_T^{\ell_{1}}$, $m^{\ell_1}_T$, and $m^{\gamma_1}_T$, the most relevant variables according to the ML algorithm. The points in orange correspond to background events, and the green ones to signal events (notice that background and signal events may overlap as can be seen from the distributions in the diagonal).
We depict the cuts corresponding to the SCB algorithm as red solid boxes. The points outside the boxes are rejected and not included in the SCB significance computation. Regarding the correlations between the variables, we can see that the SCB has not established any lower cut for $m^{\gamma_1}_T$, even when one could assume that setting $m^{\gamma_1}_T \gtrsim 75$ GeV would be preferred by analyzing only the $m^{\gamma_1}_T$ distribution (bottom-right panel of Figure~\ref{fig:relevant-variables}) dismissing its impact on the other features.

Concerning the ML-based approach, in Figure~\ref{fig:ImpVars_correlation} the contour plots represent the average of the ML output of all the events in 2D bins, which we denote $\langle o(x)\rangle$. To compute this magnitude, we average over events with the same values of the two considered features while the other features are not fixed and cover the entire analyzed ranges. If the color is the darkest (black), an event in that region has $\langle o(x) \rangle \sim 0$, meaning that the algorithm determines that the event is background-like, whereas the lightest color (white) is associated with $\langle o(x) \rangle \sim 1$, implying that the event is more likely to be classified as signal-like.
We highlight again that we are considering the average output, as the prediction for an event depends on all the variables. Nonetheless, this representation is useful to provide an idea of the behavior of the ML taking into account some correlations that otherwise are difficult to plot in a large parameter space. From Figure~\ref{fig:ImpVars_correlation} we can see that the ML algorithm performs well identifying the events not only inside the signal-enriched regions obtained with the SCB process but also in the tails of the distributions. In this case, the correlations between variables are described better, for example in the $[p_T^{\ell_{1}}, E_T^\text{miss}/\sqrt{H_T}]$ plane the rectangular cuts do not seem to be optimal. Importantly, since the ML-based methods do not need to define a region to maximize the signal-to-background ratio, they can take advantage of the entire parameter space including the events hard to classify, with $\langle o(x) \rangle \sim 0.5$, as well as the tails of the distributions. 

\begin{figure}
    \centering
    \includegraphics[width=0.49\textwidth]{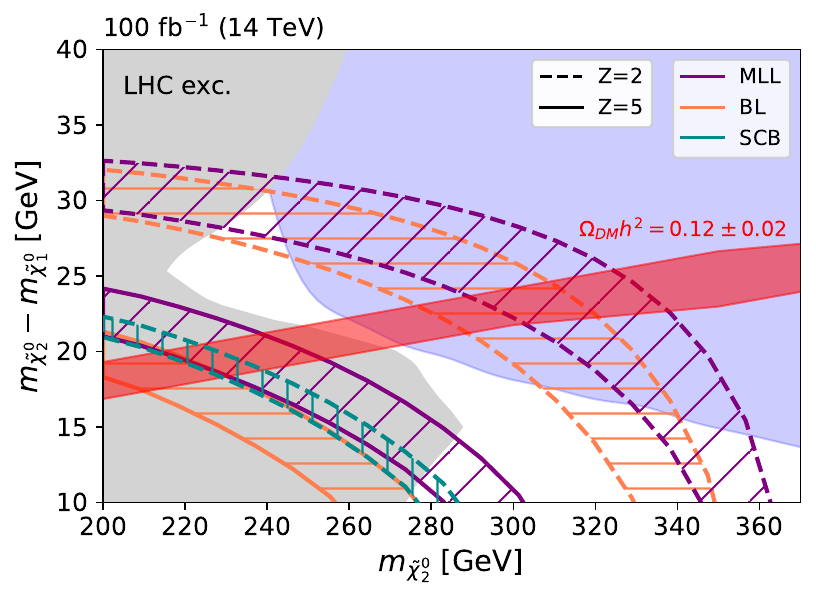}
    \caption{Projected exclusion ($Z$ = 2, dashed contour lines) and discovery reach  ($Z$ = 5, solid contour lines) in the [$m_{\tilde\chi_2^0}$, $m_{\tilde\chi_2^0}-m_{\tilde\chi_1^0}$] plane applying the optimized sequential cut-based strategy (SCB, green), the Binned-Likelihood approach (BL, orange), and the Machine-Learned Likelihood method (MLL, violet). All points in this parameter space lead to values of $\Delta a_\mu$ close to $2 \times 10^{-9}$ (see Table 1). Direct detection constraints are shown as a light-blue region, including the recent LUX-ZEPLIN results~\cite{LZCollaboration:2024lux}. LHC bounds~\cite{Chakraborti:2024pdn} are shown as a gray-shaded region. The red band indicates where the model parameters explain the observed DM relic density.}
    \label{fig:contour_plots}
\end{figure}

Our main results are summarized in Figure~\ref{fig:contour_plots} where we show the projected exclusion and discovery reach, $Z$ = 2 (dashed lines) and $Z$ = 5 (solid lines), respectively, in the [$m_{\tilde\chi_2^0}$, $m_{\tilde\chi_2^0}-m_{\tilde\chi_1^0}$] plane, for the three strategies considered in this work. 
To obtain the contour levels, we have interpolated the significance values obtained with the 15 BPs. To assess the impact of including statistical uncertainties, we have adopted a conservative approach in which we have computed how the $Z=2$ and $Z=5$ regions are weakened considering $1\sigma_{\text{stat}}$, shown as hatched regions.~\footnote{We have employed 2k pseudo-experiments for each BP to evaluate the test statistic and compute the significance, while its statistical uncertainty can be obtained from the variance of the resulting distribution. We have checked that increasing the amount of pseudo-experiments does not significantly modify the results.} To start, we find an improvement factor of $\sim 3-6$ in the discovery reach compared with the baseline significances reported in Table~\ref{tab:numberofevents}. For a fix $\tilde\chi_2^0$ mass, the sensitivity increases for smaller mass splittings. This behavior can be explained by the fact that smaller mass differences between the neutralinos result in larger branching ratios, leading to a higher number of expected signal events, as shown in Tables~\ref{table:bps} and~\ref{tab:numberofevents}. In addition, the machine-learning strategies are significantly more sensitive than the cut-based one, with the unbinned MLL method providing a slightly larger discovery reach than the BL procedure. This improvement can be illustrated by noticing that for the SCB strategy, only the $Z=2$ level is within the parameter space of interest and in the same region as the $Z=5$ reach provided by ML methods.

As can be seen in Table~\ref{table:bps}, the entire parameter space reproduces the value of the muon $g_\mu - 2$ within $2\sigma$. The models where the neutralinos reproduce the observed DM abundance is shown as a red band, while points below (above) produce DM under (over) abundance, but may be viable by additional SM modifications without affecting the collider signatures. LHC exclusion bounds taken from Ref.~\cite{Chakraborti:2024pdn} are shown in gray and covers almost all the region with $Z\ge 5$, but a small space with $275 \le m_{\tilde\chi_2^0} \le 305$ GeV and $10 \le m_{\tilde\chi_2^0}-m_{\tilde\chi_1^0} \le 15$ GeV. Direct detection limits including the recently announced LUX-ZEPLIN results~\cite{LZCollaboration:2024lux} are shown in light-blue, and are complementary to the LHC bounds excluding important regions with high neutralino masses and mass differences. We would like to highlight that the new direct detection results are almost an impressive one order of magnitude stronger than previous constraints that only excluded the region above $m_{\tilde\chi_2^0}-m_{\tilde\chi_1^0} \simeq 40$ GeV. In Appendix~\ref{app:trigger2} we show the results considering a more stringent selection criteria consistent with current LHC triggers, specifically, a cut on the missing transverse energy of $E_T^\text{miss}>$ 200 GeV. It can be seen that both results are compatible, but the latter suffer from larger statistical uncertainties due to a lower number of events.

Remarkably, almost all the allowed parameter space in Figure~\ref{fig:contour_plots} can be probed with $Z \ge 2$ by the machine-learning methods described in this article, and can be divide in three regions:
\begin{itemize}
    \item A region with $275 \le m_{\tilde\chi_2^0} \le 365$ GeV and $m_{\tilde\chi_2^0}-m_{\tilde\chi_1^0} \le 20$ GeV, where an additional DM component besides the lightest neutralino would be needed.
    \item A small region where the lightest neutralino can reproduce the right amount of dark matter abundance (red band) with $250 \le m_{\tilde\chi_2^0} \le 280$ GeV and $20 \le m_{\tilde\chi_2^0}-m_{\tilde\chi_1^0} \le 22$ GeV. This region is particularly interesting, not only because these points could satisfy all experimental constraints considering only MSSM processes, provide an explanation for the entire DM budget within the standard freeze-out paradigm and a potential solution to the small multilepton excesses~\cite{ATLAS:2019lng, ATLAS:2021moa, CMS:2021edw}, but also because it could be probed by complementary strategies, both LHC searches and direct dark matter experiments, in the near future.
    \item A region where one needs to invoke a dilution process to sufficiently reduce the DM overdensity in the mass range $m_{\tilde\chi_2^0}\sim220-250$ GeV and $22 \le m_{\tilde\chi_2^0}-m_{\tilde\chi_1^0} \le 32$ GeV, but could also explain the ATLAS and CMS collaboration multilepton excesses.
\end{itemize}

It is worth noting that the direct detection constraint are applied in the vanilla freeze-out scenario with standard cosmology if $\Omega_{\tilde \chi_1^0} h^2 \le 0.12$. As mentioned above, for models with DM overdensity we assume a dilution factor by additional SM modifications, such as a late injection of entropy during the Big Bang or the presence of another species with which the $\tilde \chi_1^0$ can coannihilate, to obtain $\Omega_{\tilde \chi_1^0} h^2 = 0.12$. However, these limits may be relaxed even further in scenarios where the dilution process leads to less than the observed relic density, thereby weakening direct detection constraints and opening the entire light-blue region of Figure~\ref{fig:contour_plots}. In that scenario, the proposed machine-learning methods can explore up to $m_{\tilde\chi_2^0} \le 365$ GeV and $m_{\tilde\chi_2^0}-m_{\tilde\chi_1^0} \le 33$ GeV. On the other hand, collider constraints are more robust since they are independent of the particular processes assumed during the early universe.

Finally, a few comments are relevant to the discussion on the potential sensitivity presented in this work. 
In our analysis, we have not included systematic uncertainties that, for example, could be incorporated as variations in the dataset after their modeling using control regions (see Appendix~\ref{app:trigger2} for an estimation of the impact of systematic uncertainties on the calculation of our limits). Therefore, we expect that the sensitivity would be reduced taking into account these effects. Nonetheless, the potential coverage of the parameter space presented in this proof-of-concept study is quite impressive and highlights the power of ML techniques.
We expect our study to encourage the experimental collaborations to perform a complete and more detailed analysis. We find this novel channel to be very promising and capable of complementing or even improving on the results of the usual di-lepton and tri-lepton plus missing energy searches~\cite{ATLAS:2019lng,CMS:2021edw,ATLAS:2021moa,CMS:2024gyw} to shed light on the challenging, yet well motivated, compressed mass scenario.

\section{Conclusions}\label{sec:concl}

The search for weakly interacting massive particles at the LHC is well motivated and, due to their small production cross section, will highly benefit from the large luminosities expected at the LHC in the coming years. Additionally, new tools made available by machine learning have allowed for more refined and precise approaches to separating signal events from large Standard Model backgrounds. A particular case of interest is when new weakly interacting massive particles decay to a potential dark matter candidate, whose experimental signature involves significant amounts of missing energy. Traditional current searches at the LHC in the framework of Supersymmetric theories include those associated with di-lepton and trilepton signatures plus missing energy and are very efficient. Recent theoretical studies, however, have proposed a new search channel, where a supersymmetric WIMP has a significant radiative decay  to the dark matter candidate. In this work we performed a first study of this new search channel and show that machine-learning techniques are promising and may complement or even improve on the results of traditional search strategies in the search for dark matter.

Our study explores electroweakino searches at the LHC, particularly for models that produce interesting but soft final states. There exists such a region in the MSSM, the ``compressed region,'' which contains both a dark matter candidate, that reproduces the observed DM relic density, and yields a contribution to $g_\mu - 2$ which can relieve the apparent tension between SM predictions and experimental observations. This region favors negative values of $\mu \times M_1$, leading to the suppression of direct DM detection cross sections, making the DM candidate a possible target for direct DM detection experiments in the coming years. This region also favors   positive values of $\mu \times M_2$ and not too heavy slepton masses (below a TeV). Combining these conditions it  happens that the second lightest neutralino tends to have a significant radiative decay  into the
lightest neutralino and a photon. 
Traditional searches, focus on multilepton final states and have placed limits on electroweakino searches in the compressed region. Here, however, we concentrate in the interesting, but yet  unused signature of  the radiative decay of the second neutralino ($\tilde \chi_2^0 \to \tilde \chi_1^0 + \gamma$). Due to the low energy of the resulting photon, cut-and-count methods typically fail to precisely carve out the kinematic region where the signal is strongest. 

In this article, we showed that the significance of electroweakino searches in the compressed region may be greatly improved employing machine-learning methods that explore the full correlation between the kinematic variables of these processes. Our analysis shows significant potential for the search of weak scale particles in this radiative decay channel, since it would allow to access most of the parameter space yet untested by other existing searches.
If neutralinos reproduces the observed DM abundance, ML techniques could probe ($Z \ge 2$) the allow region between current LHC and direct detection constraints, $250 \le m_{\tilde\chi_2^0} \le 280$ GeV and $20 \le m_{\tilde\chi_2^0}-m_{\tilde\chi_1^0} \le 22$ GeV, covering the MSSM parameter space consistent with radiative neutralino decays. If one allows additional processes which would not affect collider signatures, such as late time entropy injection or the presence of more DM candidate to dilute or complement the lightest neutralino relic production, machine-learning methods can explore up to $m_{\tilde\chi_2^0} \le 365$ GeV and $m_{\tilde\chi_2^0}-m_{\tilde\chi_1^0} \le 33$ GeV, including the entire region where ATLAS and CMS have found small multilepton excesses.
The analysis performed in this work ignores systematic errors and must be therefore corroborated by a full experimental analysis. The LHC experimental collaborations should be encouraged to proceed with such a study, which is complementary to the traditional di-lepton and tri-lepton plus missing energy ones.  

\vspace{2.5mm}
\paragraph{Acknowledgments.}

\par

The authors thank Ana Cueto G\'omez, Fernando Monticelli, and Hern\'an Wahlberg for useful discussions on systematic uncertainties, background and triggers. MC, DR, and CW would like to thank Sebastian Baum, Tong Ou, and Nausheen Shah for their collaboration in the early stages of this analysis, and Sebastian Wagner-Carena for insightful comments on our work. 
This work is partially supported by the Spanish Research Agency (Agencia Estatal de Investigaci\'on) through the grants IFT Centro de Excelencia Severo Ochoa No CEX2020-001007-S (EA, MdlR, AP, RMSS), PID2021-124704NB-I00 (EA, RMSS), RYC-2017-22986 (RMSS), CNS2023-144536 (RMSS), and PID2021-125331NB-I00 (MdlR, AP), funded by MCIN/AEI/10.13039/501100011033, and by the Comunidad Aut\'onoma de Madrid through the grant SI2/PBG/2020-00005 (MdlR, AP).
MdlR is supported by the Next Generation EU program, in the context of the National Recovery and Resilience Plan, Investment PE1 – Project FAIR ``Future Artificial Intelligence Research''.
MC\ and CW\ would like to thank the Aspen Center for Physics, which is supported by National Science Foundation grant No.~PHY-1607611, where part of this work has been done.
Fermilab is operated by Fermi Research Alliance, LLC under Contract No. DE-AC02-07CH11359 with the U.S. Department of Energy. The work of CW at the University of Chicago is also supported by the DOE grant DE-SC0013642. Work at Argonne is partially financed by the U.S. Department of Energy (DOE), Div. of HEP, Contract DE-AC02-06CH11357. This work is partially supported by the DOE under Task TeV of contract DE-FGO2-96-ER40956. This work is supported by the U.S. Department of Energy, Office of Science, Office of Workforce Development for Teachers and Scientists, and Office of Science Graduate Student Research (SCGSR) program. The SCGSR program is administered by the Oak Ridge Institute for Science and Education for the DOE under contract number DE‐SC0014664.

\section*{Appendix}
\appendix


\section{Kinematic Distributions}
\label{app:complementaryplots}

\begin{figure}
    \centering
    \includegraphics[width=1\textwidth]{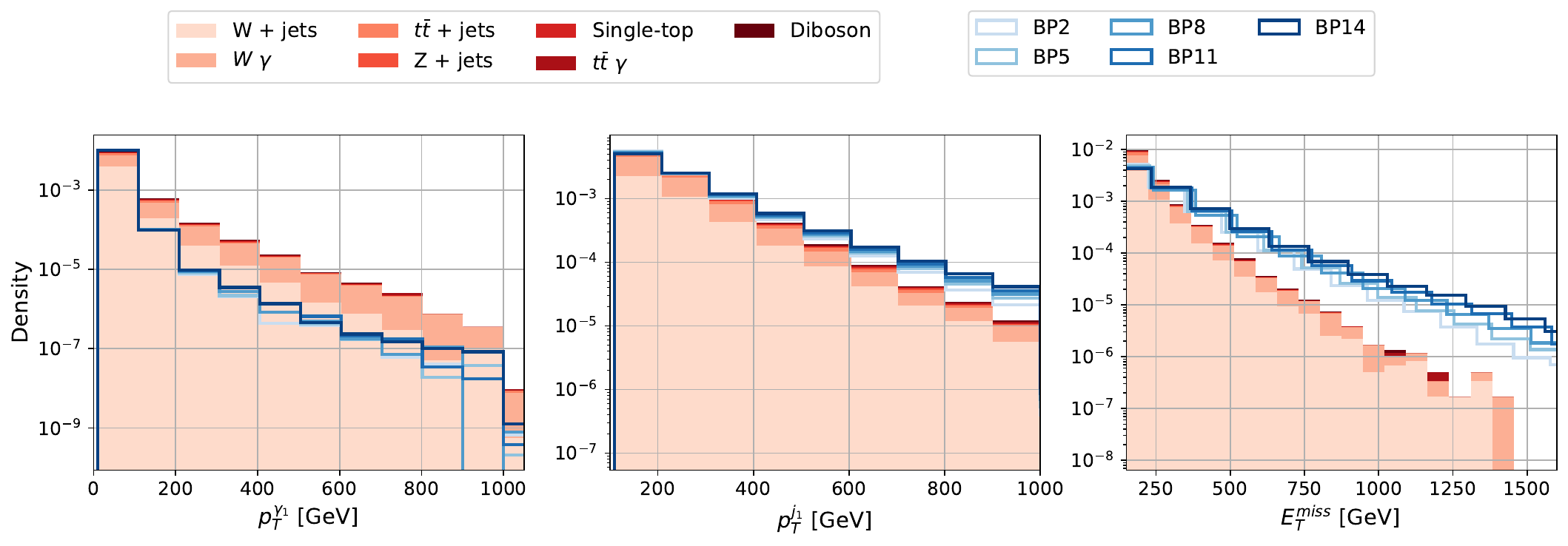}
    \includegraphics[width=1\textwidth]{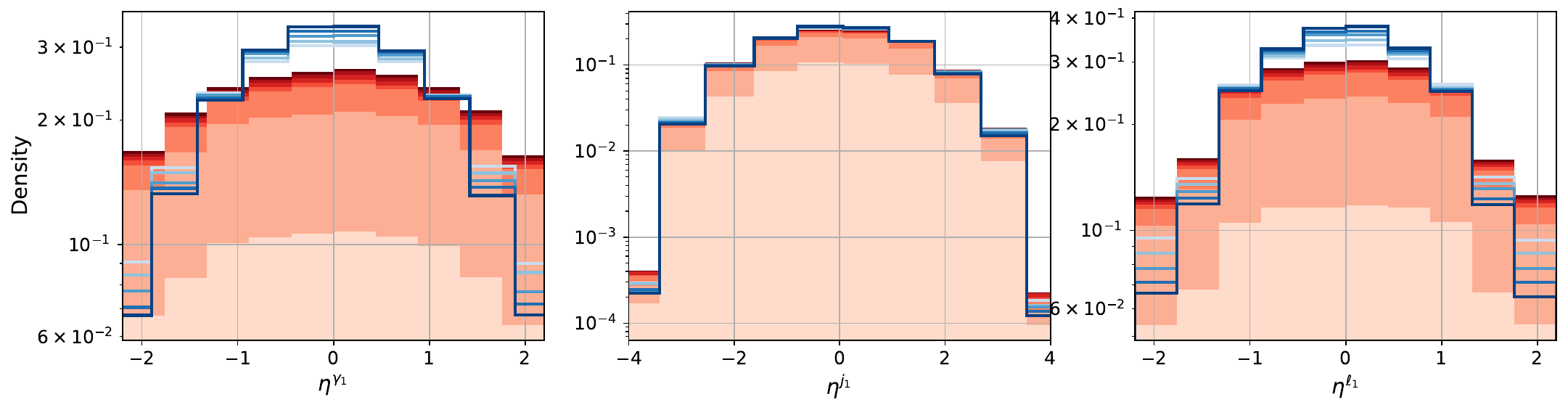}
    \includegraphics[width=1\textwidth]{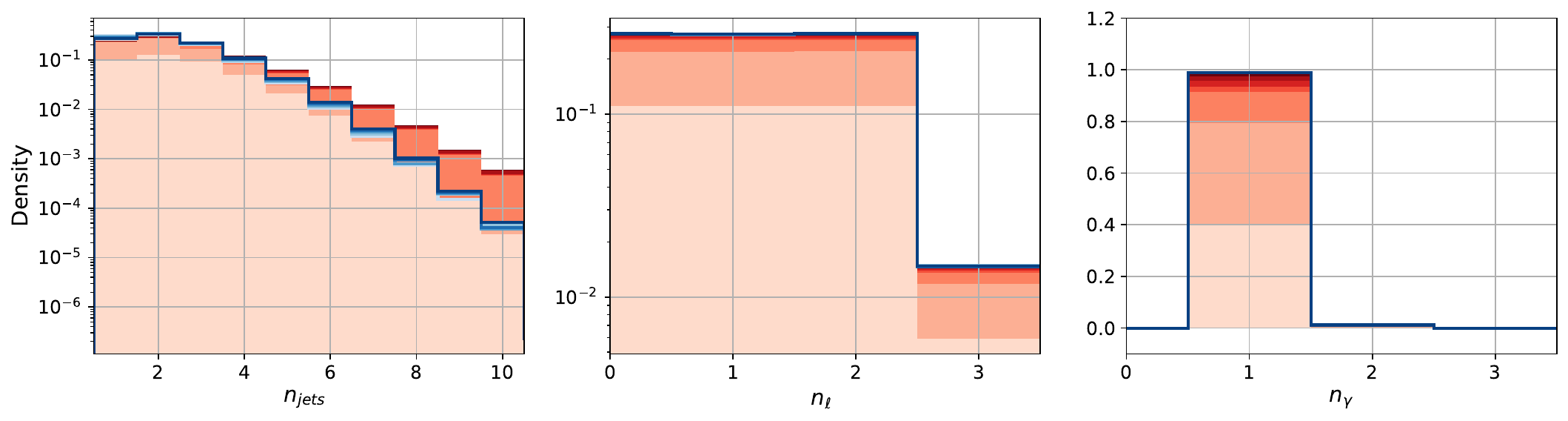}
    \caption{Total transverse momentum of the leading photon ($p_T^{\gamma_1}$) and the leading jet ($p_T^{j_1}$), missing transverse energy ($E_T^\text{miss}$), pseudorapidity ($\eta$) of the three leading visible objects in the final state ($\{\gamma_1, j_1, \ell_1 \}$) and object multiplicities, after applying event selection criteria.}
    \label{fig:low-level-app}
\end{figure}

In this appendix, we present the distributions of the features employed to analyze the proposed signal (the four most relevant variables according to the ML classifier are shown in Figure~\ref{fig:relevant-variables}).

In Figure~\ref{fig:low-level-app} we show the $p_T$ distributions for the leading photon and the leading jet, the missing transverse energy $E_{T}^{\rm{miss}}$ (upper row), the pseudorapidity $\eta$ for all the leading visible objects (middle row), as well as the object multiplicities (bottom row), both for the background and five different benchmark points leading to the correct DM relic density.

Since the characteristic feature of the proposed signal is the presence of a soft photon, the naive approach would be to focus and establish dedicated cuts on the transverse momentum of the leading photon ($p_T^{\gamma_{1}}$). However, by comparison with the kinematic variables shown in Figure~\ref{XGBoost-outputs},
it turns out to be the fifth most important feature for ML discrimination. The ISR boost leads to an increase of the kinematics of the events, and consequently of the radiative decay, increasing $p_T^{\gamma_{1}}$. On the other hand, the background photons tend to be much more energetic than the signal ones and, similarly to the lepton case, this is caused by the fact that the photon comes from the radiative decay of the wino-like second lightest neutralino $\tilde \chi_2^0$, with small mass splitting with the bino-like lightest neutralino $\tilde \chi_1^0$. Moreover, the $E_T^\text{miss}$ is much larger in the signal than in the background, as expected by the production of a pair of $\tilde \chi_1^0$ in the former, while in the background it is only generated by the presence of neutrinos from the leptonic $W$-boson decays, or purely instrumental by the presence of soft jets and the defects in the reconstruction of the events. In the case of $p_T^{j^1}$, the distribution is slightly softer for the dominant backgrounds since the jet is recoiling with objects that are lighter than in the signal case.

Regarding angular distributions, the leading photon and leading lepton are more centered in the $\eta$ distributions because they are decay products from the electroweakino-pair system, which is generated back-to-back with respect to the ISR jet, while for the background the pseudorapidity distributions are more homogeneous. In the case of the leading jet, no angular difference is observed between background and signal distributions as expected.

Regarding object multiplicities, we observe a higher $n_\text{jets}$ spectrum in the background. The dominant contribution to this effect is the $t\Bar{t}+$jets channel, coming from the hadronic decays of the $t\Bar{t}$ system (which provides an extra $W$ boson compared to the signal and the other backgrounds). 

\begin{figure}
    \centering
    \includegraphics[width=1\textwidth]{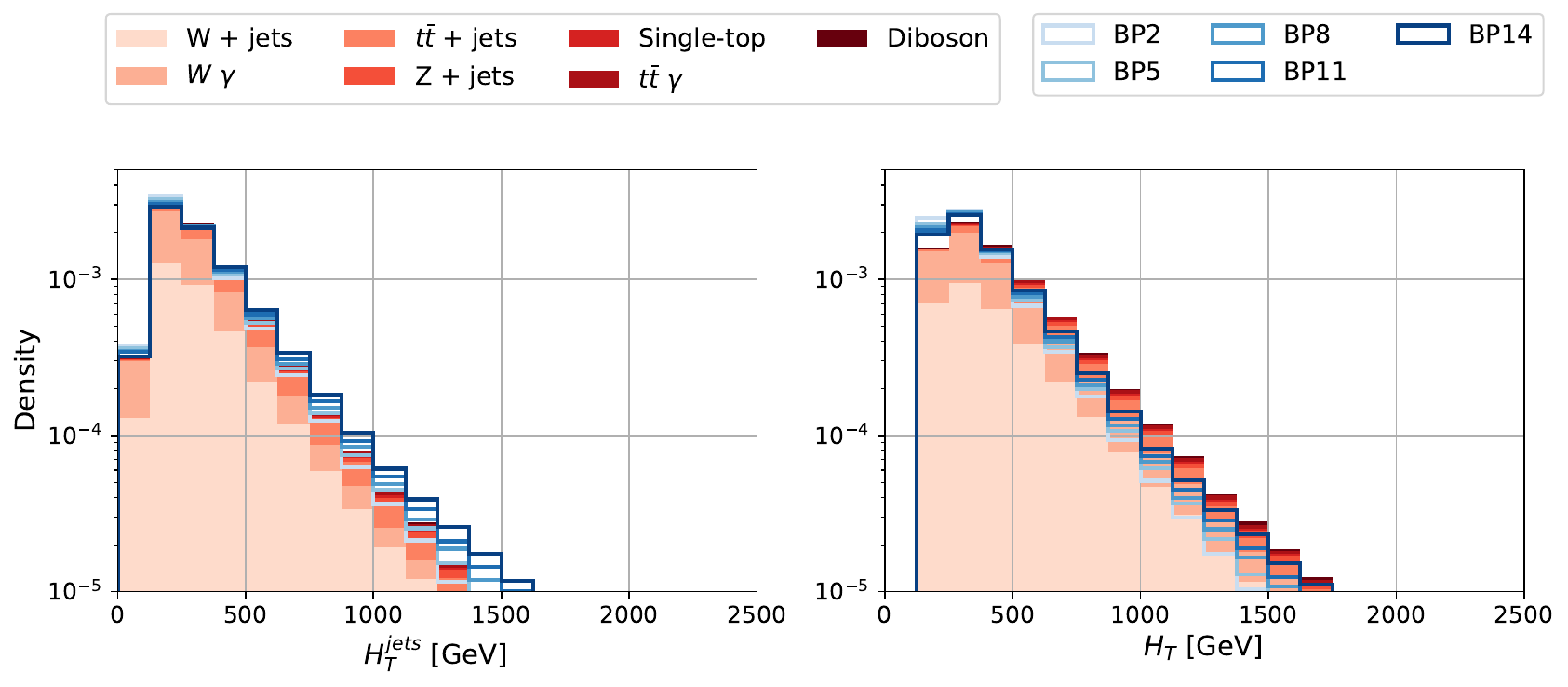}
    \includegraphics[width=1\textwidth]{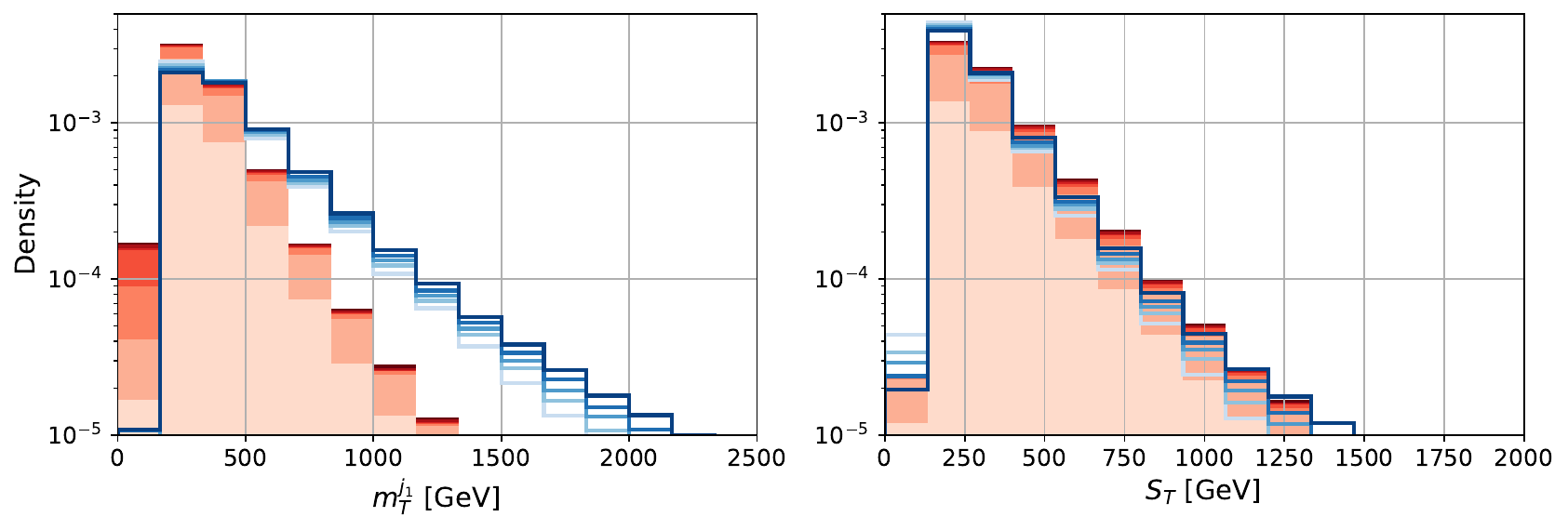}
    \caption{Hadronic activity ($H_T^{\text{jets}}$), total transverse energy ($H_T$), transverse mass of the leading jet ($m_T^{j_1}$), and scalar sum of the transverse momentum of all the leading particles ($s_T$) in the final state, after applying event selection criteria.}
    \label{fig:high-level-app}
\end{figure}

An overall insight into the behavior of the rest of high-level objects can be obtained in Figure~\ref{fig:high-level-app}. Among them, it is worth mentioning the invariant mass of the leading jet ($m_T^{j_1}$), which is higher for signal events. This behavior is expected given the size of the initial boost (higher than 100 GeV), which favors the back-to-back configuration between the leading jet and the reconstructed $E_T^{\text{miss}}$. 


\section{Review on MLL and BL Formulae}
\label{app:blmll}

In this appendix, we summarize the formulae for the MLL and the BL procedures to estimate the exclusion and discovery potential of the LHC. For the MLL method, the likelihood function $\mathcal{L}$ of $N$ independent measurements, described by an arbitrarily high-dimensional set of observations, $x$, is given by
\begin{equation}
    \mathcal{L}(\mu,s,b) = \text{Poiss}\big(N|\mu S + B\big)\,\prod_{i=1}^{N}p(x_{i}|\mu,s,b)\label{eq:stat_model} \,,
\end{equation}
where $S$ ($B$) is the expected total signal (background) yield and $\mu$ defines the hypothesis we are testing for. Additionally, Poiss stands for Poisson probability distribution, and $p(x|\mu,s,b)$ is the probability density for a single measurement $x$,
\begin{equation}
    p(x|\mu,s,b) = \frac{B}{\mu S + B}\,p_{b}(x)+\frac{\mu S}{\mu S + B}\,p_{s}(x)\,,\label{eq:prob_single_measurement}
\end{equation}
where $p_{s}(x)=p(x|s)$ ($p_{b}(x)=p(x|b)$) is the signal (background) probability density function (PDF) for $x$, encoding the event-by-event information.
The discovery reach analysis corresponds to studying the background-only hypothesis ($\mu = 0$), then the log-likelihood ratio test statistic becomes
\begin{equation}
\tilde{q}_{0} =   \begin{cases}
0 & \rm{if} \, \, \hat{\mu} < 0 \\
-2 \text{ Ln } \frac{\mathcal{L}(0,s,b)}{\mathcal{L}(\hat{\mu},s,b)} =  -2\hat{\mu} S + 2 \sum_{i=1}^{N} \text{ Ln } \left( 1 + \frac{\hat{\mu} S p_s(x_i)}{B p_b(x_i)}\right)  &  \rm{if} \, \,  \hat{\mu} \geq 0\,,
\end{cases}
  \label{eq:testdiscovery}
\end{equation}
where $\hat{\mu}$ is the parameter that maximizes the likelihood. The value of $\hat{\mu}$ can be estimated numerically by calculating the zero of the partial derivative of Eq.~\eqref{eq:stat_model} with respect to $\mu$.

Since $p_{s,b}(x)$ are unknown, the base idea used by the MLL method is to replace these densities for the one-dimensional classification score $o(x)$ that can be obtained from a machine-learning binary classifier,
\begin{equation}
    p_{s}(x) \rightarrow \tilde{p}_{s}(o(x))\,,\,\,\, p_{b}(x) \rightarrow \tilde{p}_{b}(o(x))\,,
    \label{reduction}
\end{equation}
where $\tilde{p}_{s}(o(x))$ ( $\tilde{p}_{b}(o(x))$) is the distribution of $o(x)$ for signal (background), obtained by evaluating the output of the classifier on a set of pure signal (background) events. Since the distributions are one-dimensional, they can be easily handled. The KDE method is then used to extract these distributions without binning and incorporate them into Eq.(\ref{eq:testdiscovery}).

The test statistic in Eq.~\eqref{eq:testdiscovery} is estimated through a finite data set of $N$ events, with the probability distribution conditioned on the true unknown signal strength $\mu'$.  When the true hypothesis is assumed to be the signal-plus-background ($\mu'=1$), the median expected discovery significance $\text{med }[Z_{0}| 1]$ is defined as
\begin{equation}
    Z_{\text{MLL}}=\text{med }[Z_{0}|1] = \sqrt{\text{med }[\tilde{q}_{0}|1]}\,,
\end{equation}
where we estimate the $\tilde{q}_{0}$ distribution numerically by generating a set of pseudo-experiments with signal-plus-background events. The value of $Z_{\text{MLL}}$ encodes how likely is to the background-only
hypothesis to explain data that follows the signal-plus-background hypothesis, and a higher value of $Z_{\text{MLL}}$ excludes the background-only hypothesis with higher confidence.

For the BL method, we consider the output of the ML, $o(x) \in [0,1]$, and divide its range in a histogram-based approach. We can describe the population of each bin as an independent Poissonian distribution, then likelihood function $\mathcal{L}$ is given by
\begin{equation}
    \mathcal{L}(\mu,s,b) = \prod_{i=1}^{N} \text{Poiss}\big(N_d|\mu S_d + B_d\big),
\end{equation} 
with $S_{d}$ and $B_{d}$ defined as the expected number of signal and background events in each bin $d$. We can repeat the previous procedure to obtain the log-likelihood ratio test statistic, however through the use of Asimov data sets~\cite{Cowan:2010js} the final formula for the discovery potential can be estimated as
\begin{equation}
    Z_{\text{BL}}=\text{med }[Z_{0}|1] = \sqrt{\text{med }[\tilde{q}_{0}|1]}=\left[2\sum_{d=1}^{D}\left((S_d+B_d)\text{ Ln}\left(1+\frac{S_d}{B_d}\right)-S_d\right)\right]^{1/2}\,.
    \label{binned-Z}
\end{equation}


\section{Impact of a more stringent trigger and systematic uncertainties}
\label{app:trigger2}

In this appendix, we show how the results are affected by adopting a more stringent cut on the missing transverse energy, and we provide an estimate of the impact of systematic uncertainties on our final limits.

We begin by discussing a more restrictive selection cut applied during the data processing phase. The only difference with the cuts described in the main body is that we tighten the minimum value of $E_T^\text{miss}$ from 100 GeV to 200 GeV. 
This value satisfies current LHC selection triggers for $E_T^\text{miss}$~\cite{ATL-DAQ-PUB-2019-001}. In Table~\ref{tab:numberofevents2} we show the expected number of events with the new selection cuts. 
Notice that we expect approximately half of the signal events compared with the ones in Table~\ref{tab:numberofevents} but 4 times less background events, which results in a similar $S/\sqrt{B}$.

\begin{table}[hbtp]
    \centering
    \begin{tabular}{c|c c c|c|c}
       \cline{1-2} \cline{4-6}
       Process  & Yield &  \hspace{1cm} & BP \# & Yield & $S/\sqrt{B}$ \\
       \cline{1-2} \cline{4-6}
        $W+\text{jets}$  & $ 10834$ & & 1 & $81$ & $0.53$\\
        $W\gamma$        & $ 11002$ & & 2 & $206$ & $1.37$\\
        $t\Bar{t}\gamma$ & $ 700 $  & & 3 & $344$ & $2.29$\\
        \cline{1-2}
        Total background & $ 22536$& & 4 & $49$ & $0.32$\\
        \cline{1-2}
        \multicolumn{2}{c}{\multirow{8}{1cm}{}} & & 5 & $ 112$ & $0.74$ \\
        \multicolumn{2}{c}{\multirow{8}{1cm}{}} & & 6 & $ 179$ & $1.19$ \\
        \multicolumn{2}{c}{\multirow{8}{1cm}{}} & & 7 & $ 32$  & $0.21$ \\
        \multicolumn{2}{c}{\multirow{8}{1cm}{}} & & 8 & $ 65$ & $0.43$ \\
        \multicolumn{2}{c}{\multirow{8}{1cm}{}} & & 9 & $ 97$ & $0.64$ \\
        \multicolumn{2}{c}{\multirow{8}{1cm}{}} & & 10 & $ 20$ & $0.13$ \\
        \multicolumn{2}{c}{\multirow{8}{1cm}{}} & & 11 & $ 38$ & $0.25$ \\
        \multicolumn{2}{c}{\multirow{8}{1cm}{}} & & 12 & $56$ & $0.37$ \\
        \multicolumn{2}{c}{\multirow{8}{1cm}{}} & & 13 & $ 12$ & $0.07$ \\
        \multicolumn{2}{c}{\multirow{8}{1cm}{}} & & 14 & $ 22$ & $0.14$ \\
        \multicolumn{2}{c}{\multirow{8}{1cm}{}} & & 15 & $ 32$ & $0.21$ \\
        \cline{4-6}
    \end{tabular}
    \caption{Expected background and signal events at the LHC with a center-of-mass energy of 14 TeV and a total integrated luminosity of  100 fb$^{-1}$, after applying same event selection criteria described in the main text but with $E_T^\text{miss} > 200$ GeV.}
    \label{tab:numberofevents2}
\end{table}

\begin{figure}
    \centering

    \includegraphics[width=0.49\textwidth]{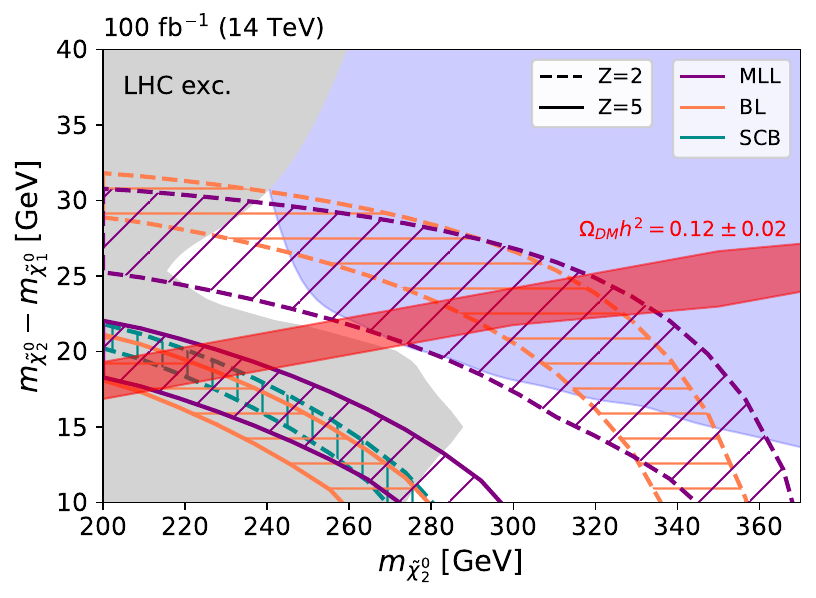}
    \includegraphics[width=0.49\textwidth]{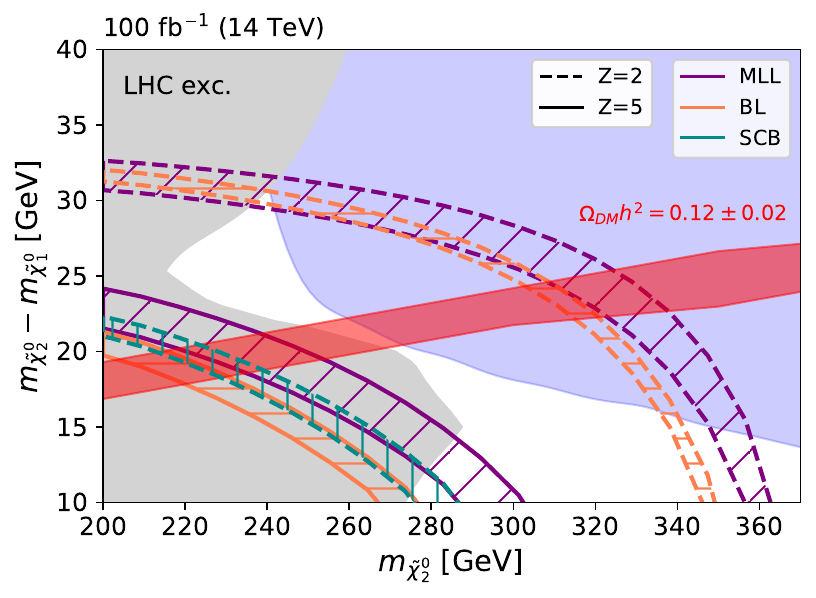}
    \caption{Left panel: projected discovery significance as shown in Fig.~\ref{fig:contour_plots} but considering the event selection criteria with $E_T^\text{miss} > 200$ GeV. Right panel: same as Fig.~\ref{fig:contour_plots} ($E_T^\text{miss} > 100$ GeV) but considering only systematic uncertainties as detailed in the text.}
    \label{fig:contour_plots2}
\end{figure}

In the left panel of Fig.~\ref{fig:contour_plots2} we present the discovery limits considering the selection criteria with $E_T^\text{miss} > 200$ GeV. Comparing the results with the ones presented in Fig.~\ref{fig:contour_plots}, we can see that the overall reach is only slightly less sensitive than those derived with $E_T^\text{miss} > 100$ GeV. Additionally, the more stringent cuts lead to fewer statistics and therefore larger statistical uncertainties as a drawback.

Next, we present an estimation of the impact of systematic uncertainties for ML-based methods that aims to be a first approximation to evaluate the stability of our results. For this, we first need to translate the systematic uncertainties in the physical-based space to the ML classifier output space. For that, in this analysis we consider 5\% systematic shifts on the most relevant variable for the ML discrimination, the missing transverse energy significance, $E_T^\text{miss}/\sqrt{H_T}$ (see the right panel of Fig.~\ref{XGBoost-outputs}).
The choice of 5\% follows from the study of the CMS detector performance of Ref.~\cite{CMS:2019ctu}, for a process with final state and selection cuts similar to the ones we explore in our analysis.

Following Refs.~\cite{Arratia:2021otl,CMS:2022ytw,Arganda:2023qni}, we assess the impact on the ML output by building two modified test data sets. These sets contain the entire set of test events with all the kinematic variables used as input variables unchanged, except for $E_T^\text{miss}/\sqrt{H_T}$ whose values are increased by a 5\% in the first new sample, denoted $TS+$, and a second sample with $E_T^\text{miss}/\sqrt{H_T}$ values decreased by the same percentage, denoted $TS-$. Then, we evaluate the original ML classifier (trained with no uncertainties) and evaluate it with both new test samples to obtain two ML outputs: $o(TS+)$ and $o(TS-)$. Finally, we repeat the computation of the significance with the binned and unbinned methods using each ML output and obtain an uncertainty band that reflects the variations produced by the considered shifts. To be conservative, we do not allow an artificial increase in significance reach, which would be a non-physical consequence due to the overestimation produced by the common increase/decrease shift of all test events and the fact that we are not imposing energy-momentum conservation.

In the right panel of Fig.~\ref{fig:contour_plots2} we show the results and, comparing with Fig.~\ref{fig:contour_plots}, we can see that the impact of systematic uncertainties is smaller than the statistical errors by a factor $\sim2$. In particular, the unbinned method is more sensitive to systematic effects than the binned and SCB approaches. This can be understood as the binning introduces a smoothing that compensates for the shift produced by systematic uncertainties if the bins are wide enough.
A precise combination of statistical and systematic errors is hence expected to be dominated by the former, and is beyond the scope of this article since can only be efficiently be performed by the CMS and ATLAS collaborations in their experimental analysis.

\bibliographystyle{JHEP} 
\bibliography{biblio.bib}

\end{document}